\documentclass{article}
\usepackage{jcappub}

\usepackage{epstopdf}
\usepackage{graphicx}
\usepackage{color}{
\usepackage[usenames,dvipsnames]{xcolor}
\usepackage{hyperref}
\usepackage[export]{adjustbox}

\def\kv{{\bf k}}
\def\qv{{\bf q}}
\newcommand{\be}{\begin{equation}}
\newcommand{\ee}{\end{equation}}
\newcommand{\bea}{\begin{eqnarray}}
\newcommand{\eea}{\end{eqnarray}}
\newcommand{\bdm}{\begin{displaymath}}
\newcommand{\edm}{\end{displaymath}}

\newcommand{\lcdm}{$\Lambda$CDM}

\newcommand{\hmone}{$\,h^{-1}$}

\newcommand{\planck}{{\it Planck}}

\def\d{\delta}
\def\lan{\langle}
\def\ran{\rangle}

\def\Mpc{\, h^{-1} \, {\rm Mpc}}
\def\kpc{\, h^{-1} \, {\rm kpc}}

\def\kMpc{\, h \, {\rm Mpc}^{-1}}

\def\ie{{\em i.e.}~}
\def\eg{{{\em e.g.}}}

\newcommand{\gadgetthree}{\textsc{gadget-3}}

\title{DEMNUni: Massive neutrinos and the bispectrum of large scale structures}

\author[a]{Rossana Ruggeri,}
\author[b]{Emanuele Castorina,}
\author[c,d,e]{Carmelita Carbone,}
\author[f,g]{Emiliano Sefusatti}

\affiliation[a]{Institute of Cosmology \& Gravitation, University of Portsmouth, Dennis Sciama Building, Portsmouth PO1 3FX --- UK}
\affiliation[b]{ Berkeley Center for Cosmological Physics, University of California, Berkeley, CA 94720, USA, and Lawrence Berkeley National Laboratory, 1 Cyclotron Road, Berkeley, CA 93720, USA}
\affiliation[c]{Universit\`a degli Studi di Milano - Dipartimento di Fisica, via Celoria 16, I-20133 Milano (MI), Italy}
\affiliation[d]{INAF - Osservatorio Astronomico di Brera, via Brera 28, I-20121 Milano (MI) -- Italy}
\affiliation[e]{INFN - Sezione di Milano, via Celoria 16, I-20133, Milano (MI) -- Italy}
\affiliation[f]{INAF - Osservatorio Astronomico di Trieste, via Tiepolo 11, I-34143 Trieste -- Italy}
\affiliation[g]{INFN - Sezione di Trieste, Via Valerio 2, I-34127 Trieste -- Italy}
\emailAdd{rossana.ruggeri@port.ac.uk}
\emailAdd{ecastorina@berkeley.edu.}
\emailAdd{carmelita.carbone@unimi.it}
\emailAdd{emiliano.sefusatti@brera.inaf.it}

\abstract{The main effect of massive neutrinos on the large-scale structure consists in a few percent suppression of  matter perturbations on all scales below their free-streaming scale. Such effect is of particular importance as it allows to constraint the value of the sum of neutrino masses from measurements of the galaxy power spectrum. In this work, we present the first measurements of the next higher-order correlation function, the bispectrum, from N-body simulations that include massive neutrinos as particles. This is the simplest statistics characterising the non-Gaussian properties of the matter and dark matter halos distributions. We investigate, in the first place, the suppression due to massive neutrinos on the matter bispectrum, comparing our measurements with the simplest perturbation theory predictions, finding the approximation of neutrinos contributing at quadratic order in perturbation theory to provide a good fit to the measurements in the simulations. On the other hand, as expected, a linear approximation for neutrino perturbations would lead to ${\mathcal O}(f_\nu)$ errors on the total matter bispectrum at large scales. We then attempt an extension of previous results on the universality of linear halo bias in neutrino cosmologies, to non-linear and non-local corrections finding consistent results with the power spectrum analysis.}


\begin{document}

\maketitle

\section{Introduction}
\label{intro}

Despite the constant improvement in the quality of observations over the last decade, the Standard, $\Lambda$CDM, Cosmological model, still provides a good fit to several cosmological observables \citep{PLANCK2016cp, AlamEtal2016a, ZhaoEtal2017a}. It is therefore clear that the analysis of future galaxy survey data \citep{LaureijsEtal2011, LeviEtal2013, LSST2017} will require accurate predictions, at the percent level or better, since any departure from the standard scenario will likely be relatively small \citep{McDonaldEisenstein2007, EisensteinSeoWhite2007, CarboneMangilliVerde2011, GiusarmaEtal2011, AmendolaEtal2013, WeinbergEtal2013, AudrenEtal2013, JoyceEtal2015, BaldaufEtal2016, DesjacquesJeongSchmidt2016, Koyama2016, AlonsoEtal2017}.

In this context, it has become necessary to properly account for the effects of neutrino masses on cosmological observables. Massive neutrinos represent a small but non negligible fraction of the matter density, characterised by a significant thermal velocity distribution even when non-relativistic, and we should account for the different evolution of their perturbations with respect to the cold dark matter component. Their overall effect consists in a damping of matter perturbations on all scales below their free-streaming scale: a several percent reduction on the {\em total} matter power spectrum, \citep{LesgourguesPastor2006, Lesgourgues2013}. This happens precisely on the range of scales probed by current galaxy redshift surveys (tens to hundreds of Megaparsecs). In fact, cosmological observations are able to provide an upper limit to the (sum of) neutrino masses \citep{PalanqueDelabrouilleEtal2015, PLANCK2016SZClusters, AlamEtal2016a, CuestaNiroVerde2016, VagnozziEtal2017, ArchidiaconoEtal2017} and, in the long run, possibly bridge the gap with the lower limit given by neutrino oscillation experiments \citep{LeviEtal2013, LaureijsEtal2011}. In other terms, neutrino masses are not simply a nuisance in the possible detection of dark energy or new physics effects, but represent an important test for the Standard Model of particles physics and its extensions. 

Several studies had appeared, over the last few years, on massive neutrinos effects on the matter power spectrum nonlinear evolution and redshift-space distortions in the context of perturbation theory  \citep{SaitoTakadaTaruya2008, Wong2008, LesgourguesEtal2009, UpadhyeEtal2014, BlasEtal2014, DupuyBernardeau2014, DupuyBernardeau2015, CastorinaEtal2015, FuhrerWong2015, UpadhyeEtal2016, ArchidiaconoHannestad2016, LeviVlah2016, LiuEtal2017A}, on baryonic acoustic oscillations \citep{VillaescusaEtal2014, LiuEtal2017}, on the halo mass function \citep{IchikiTakada2012, CastorinaEtal2014, CostanziEtal2013B, LoVerde2014B, CastorinaEtal2015}, halo bias \citep{BiagettiEtal2014, LoVerdeZaldarriaga2014, LoVerde2016, CastorinaEtal2014, CastorinaEtal2015, ChiangEtal2017} and cosmic voids \citep{MassaraEtal2015}. Novel probes of massive neutrinos have also been recently proposed \citep{ZhuEtal2014, OkoliEtal2016}. 

Focusing on the theoretical description of matter perturbations, predictions for the nonlinear matter power spectrum have been first studied in \citep{SaitoTakadaTaruya2008} and \citep{Wong2008}. The model proposed for the {\em total} matter power spectrum approximates neutrino perturbations with their linear prediction (obtained from a Boltzmann code, and therefore acting as a source for cold and baryonic matter), limiting nonlinear corrections only to the {\em cold} dark matter component. The limits of this approximation are carefully studied in both works, estimating systematic errors due to the linear neutrino assumption to be at the sub-percent level. Indeed, comparisons with particle-based N-body simulations show the discrepancies between numerical results and predictions based on Perturbation Theory (PT) in the massive neutrino case to be consistent with the typical accuracy of the standard PT approach in $\Lambda$CDM models \citep{CastorinaEtal2015}.  The validity of the linear neutrino approximation is further explored in \citep{BlasEtal2014, SenatoreZaldarriaga2017} where the authors highlight how the violation of momentum conservation inherent in the scheme might have significant effects in the nonlinear corrections beyond the 1-loop level. They propose an hybrid model combining the full Boltzmann treatment at high redshift and the two-fluid approximation at later times as a starting point for studying the nonlinear evolution. In this approximation, both cold dark matter (including baryons) and neutrinos are described as fluids, the second characterised by a (time-dependent) effective speed of sound, estimated from the neutrino velocity dispersion \citep{ShojiKomatsu2010}. Explicit predictions for the matter bispectrum have been presented, so far, in \citep{FuhrerWong2015} and \citep{LeviVlah2016}. Ref. \citep{FuhrerWong2015} presents a test of the two-fluid and the linear neutrinos approximations, against the exact treatment via the collisionless Boltzmann equation, using the bispectrum as a specific measure of their validity at the level of higher-order corrections. They show how both approximations fail to provide a 1\% accuracy on the total matter bispectrum. In particular, as we will discuss, the linear neutrinos approximation does not correctly reproduce the large scale matter bispectrum for large neutrino masses, while in the limit of a small neutrino density fraction, $f_\nu$, becomes significantly more accurate.  The same limit is also explored in the alternative formulation of \citep{LeviVlah2016}, where a perturbative expansion around $f_\nu=0$ is considered. 

In this work we present, for the first time, measurements of the matter and halo bispectrum in N-body simulations that include a massive neutrino component as particles. In the case of the two matter components, neutrinos and cold dark matter/baryons, we provide a first comparison with theoretical predictions assuming neutrinos perturbations at next-to-leading order in PT. In addition,  we attempt an extension of the results of \citep{CastorinaEtal2014} on the ``universality'' of linear halo bias to the nonlinear level, including as well non-local corrections \citep{ChanScoccimarroSheth2012, BaldaufEtal2012}. In fact, the bispectrum, as a direct result of nonlinear evolution, provides a valuable test of nonlinear effects (due to both gravitational instability and bias) in these, by now, standard cosmological models. Moreover, the galaxy bispectrum, is the lowest-order statistic encoding and quantifying the non-Gaussian properties of the galaxy distribution \citep{SefusattiEtal2006}. Several groups have measured three-point statistics in recent data \citep{SlepianEtal2016, GilMarinEtal2017}, showing that the adding this information to the standard power spectrum analysis yield to better constraints on the cosmological parameters. 

The paper is organized as follow. In Sec.~\ref{sec:linear_nu} we briefly review basic results on massive neutrinos perturbations. In Sec.~\ref{sec:bispm} we show our measurements of the bispectrum for the various matter components and compare them with theoretical predictions from perturbation theory, while in Sec.~\ref{sec:bisph} we show the halo bispectrum measurements and derive the corresponding halo bias functions.  In Sec.~\ref{sec:conclusions} we conclude summarising  the results obtained.

\section{Cosmological perturbations in the presence of massive neutrinos}
\label{sec:linear_nu}

\subsection{Linear evolution}

In the early Universe neutrinos are kept in equilibrium with baryons and photons via weak interactions, eventually decoupling at a temperature of about $T_{\rm dec} \sim 9\times 10^9\;\rm K$, when their interaction rate becomes comparable to rate of cosmological expansion. After decoupling, neutrinos free-stream with large thermal velocities described by a Fermi-Dirac distribution. As the Universe expands neutrinos slow down, becoming non relativistic at a typical redshift of 
\be
1+z_{nr} \simeq 1980\,\frac{m_{\nu,i}}{1 \text{ eV}}\,,
\ee
where $m_{\nu,i}$ represents the neutrino mass eigenstate $i$ in electronvolt. As non-relativistic particles, while contributing to the total matter density the fraction
\be
f_\nu\equiv\frac{\Omega_\nu}{\Omega_m}=\frac{1}{\Omega_{m,0}h^2}\frac{\sum_i\,m_{\nu,i}}{93.14~{\rm eV}}\,,
\ee
they still travel much longer distances,of the order of tens of Megaparsecs, compared to standard cold dark matter (CDM) particles. We define a free-streaming length $\lambda_{\rm fs} \propto v_{th}/ H(t)$, with $v_{th}$ being the characteristic thermal velocity of neutrino particles. Below this scale we expect a suppression of the neutrino density fluctuations with respect to CDM ones. On scales $\lambda\gg\lambda_{\rm fs}$ we expect neutrinos to behave as CDM. In Fourier-space, the wavenumber corresponding to $\lambda_{\rm fs}$ can be written as
\be
k_{\rm fs}\simeq\frac{0.908}{\sqrt{1+z}}\,\sqrt{\Omega_{m,0}}\,\frac{m_{\nu,i}}{1~{\rm eV}}\kMpc\,.
\ee

For simplicity we denote with the subscript ``{\em c}'' quantities related to the CDM {\em and} baryonic matter components as no distinction will be made between the two.  We refer to such component generically as ``cold'' matter, as opposed to the neutrinos contribution. Total matter perturbations are therefore given by the weighted sum
\be\label{eq:deltam}
\delta_m =(1- f_{\nu})\delta_c +f_{\nu} \delta_{\nu}\, .
\ee
with $ \delta_{\nu}$ denoting neutrinos perturbations. 

On scales smaller than $\lambda_{\rm fs}$ neutrinos do not provide support to the Newtonian gravitational potential, and we expect the growth of cold matter perturbations to be different with respect to a standard cosmology. It is possible to show \citep{HuEisensteinTegmark1998} that, in linear theory, for $k\gg k_{\rm fs}$, assuming a constant $\Omega_{m}$ and a constant amplitude for primordial fluctuations, the ratio of the cold matter power spectra in a cosmology with $f_\nu\ne0$ to the massless neutrino case is given by 
\be\label{eq:suppPcc}
\frac{P_{cc}^{f_\nu \neq 0 }(k)  }{P_{cc}^{f_\nu = 0 } (k) }\, \stackrel{k\gg k_{\rm fs}}{ \simeq}\, 1- 6\, f_{\nu}. 
\ee
The total matter power spectrum, from eq.~\ref{eq:deltam}, can be written as 
\be\label{eq:Pmm}
 P_{mm}(k) = (1-f_{\nu})^2\,P_{cc}(k) + 2f_{\nu}(1-f_{\nu})P_{\nu c}(k)+ f_{\nu}^2 P_{\nu\nu}(k) \,,
\ee
with $P_{\nu c}(k)$ denoting the cold matter-neutrinos cross-power spectrum. Neglecting neutrino perturbations at small scales we obtain the well-known limit \citep{HuEisensteinTegmark1998}
\be\label{eq:suppPmm}
\frac{P_{mm}^{f_\nu \neq 0 }(k)  }{P_{mm}^{f_\nu = 0 } (k) }\, \stackrel{k\gg k_{\rm fs}}{ \simeq}\, 1- 8 ~f_{\nu}\,. 
\ee

\subsection{Nonlinear evolution}

Numerical simulations \citep{BrandbygeEtal2008, VielHaehneltSpringel2010, BirdVielHaehnelt2012, WagnerVerdeJimenez2012, InmanEtal2015,YuEtal2017} have shown that the suppression expected in linear theory according to eq.~\ref{eq:suppPcc} and \ref{eq:suppPmm} is enhanced at the nonlinear level, as expected from predictions in perturbation theory \citep{SaitoTakadaTaruya2008, Wong2008}, with important consequences for the constraints on neutrino masses \citep{SaitoTakadaTaruya2008, AudrenEtal2013}. 

On the other hand, a direct consequence of the nonlinear evolution of matter fluctuations is given by the emergence of non-Gaussianity, quantified, in the first place by a non-vanishing matter bispectrum. Denoting the cold matter fraction as $f_c\equiv 1- f_\nu$, the analogue of eq.~\ref{eq:Pmm} for the total matter bispectrum is given by 
\bea\label{eq:Bmmm}
B_{mmm}(k_1,k_2,k_3) & = & f^3_c\,B_{ccc}(k_1,k_2,k_3)+f_c^2\,f_\nu\,B_{cc\nu}^{(s)}(k_1,k_2,k_3)\nonumber \\
& & + f_c\,f_\nu^2\,B_{\nu\nu c}^{(s)}(k_1,k_2,k_3)+ f_\nu^3\,B_{\nu\nu\nu}(k_1,k_2,k_3) 
\eea
where we introduced the symmetrized versions of the cross cold-cold-neutrino bispectrum $B_{cc\nu}^{(s)}$ and neutrino-neutrino-cold bispectrum $B_{\nu\nu c}^{(s)}$ defined, respectively, as  
\bea\label{eq:Bccn}
\d_D(\kv_{123})\, B_{cc\nu}^{(s)}(k_1,k_2,k_3) & \equiv &
\left[\lan \d_c(\kv_1)\d_c(\kv_2)\d_\nu(\kv_3)\ran\right.\nonumber\\
& & \left.+\lan \d_c(\kv_3)\d_c(\kv_1)\d_\nu(\kv_2)\ran+\lan \d_c(\kv_2)\d_c(\kv_3)\d_\nu(\kv_1)\ran\right]
\eea
and 
\bea\label{eq:Bnnc}
\d_D(\kv_{123})\, B_{\nu\nu c}^{(s)}(k_1,k_2,k_3) & \equiv &
\left[\lan \d_c(\kv_1)\d_\nu(\kv_2)\d_\nu(\kv_3)\ran\right.\nonumber\\
& & \left.+\lan \d_c(\kv_3)\d_\nu(\kv_1)\d_\nu(\kv_2)\ran+\lan \d_c(\kv_2)\d_\nu(\kv_3)\d_\nu(\kv_1)\ran\right]\,.
\eea
where we made use of the notation $k_{ij}\equiv\kv_i+\kv_j$ for vectors sums.

Although neutrinos cluster very weakly below the free streaming scale, on larger scales they behave like CDM. This means that assuming {\em linear} neutrino perturbation on {\em any} scale results in a lack of power on the large-scale bispectrum, which is the outcome of the nonlinear evolution of all matter components. As a consequence, numerical simulations treating neutrinos only at the linear level on the grid (or not having neutrino fluctuations at all) will predict the wrong bispectrum as well as any other higher-order correlation function. 
On the other hand the two fluids behave similarly only in the very low-$k$ regime, and therefore we expect next -to-leading order, \ie quadratic, correction to capture most of the neutrino contributions to the bispectrum terms in the above equations.  
In analogy with the work in  \citep{SaitoTakadaTaruya2008, CastorinaEtal2015} for the matter power spectrum in cosmologies with massive neutrinos we assume neutrinos to contribute only at tree-level in the PT calculation, whereas we compute CDM density perturbations up to the 1-loop level.
The analysis in \cite{CastorinaEtal2015} showed that this simple approach reproduces the measurements of the power spectrum in simulations to 1\% accuracy. The goal of this section is to test on simulations whether the same assumptions hold true for the bispectrum, within the limits of the precision of our measurements.

\begin{figure}[t]
\begin{center}
\includegraphics[width=.45\textwidth]{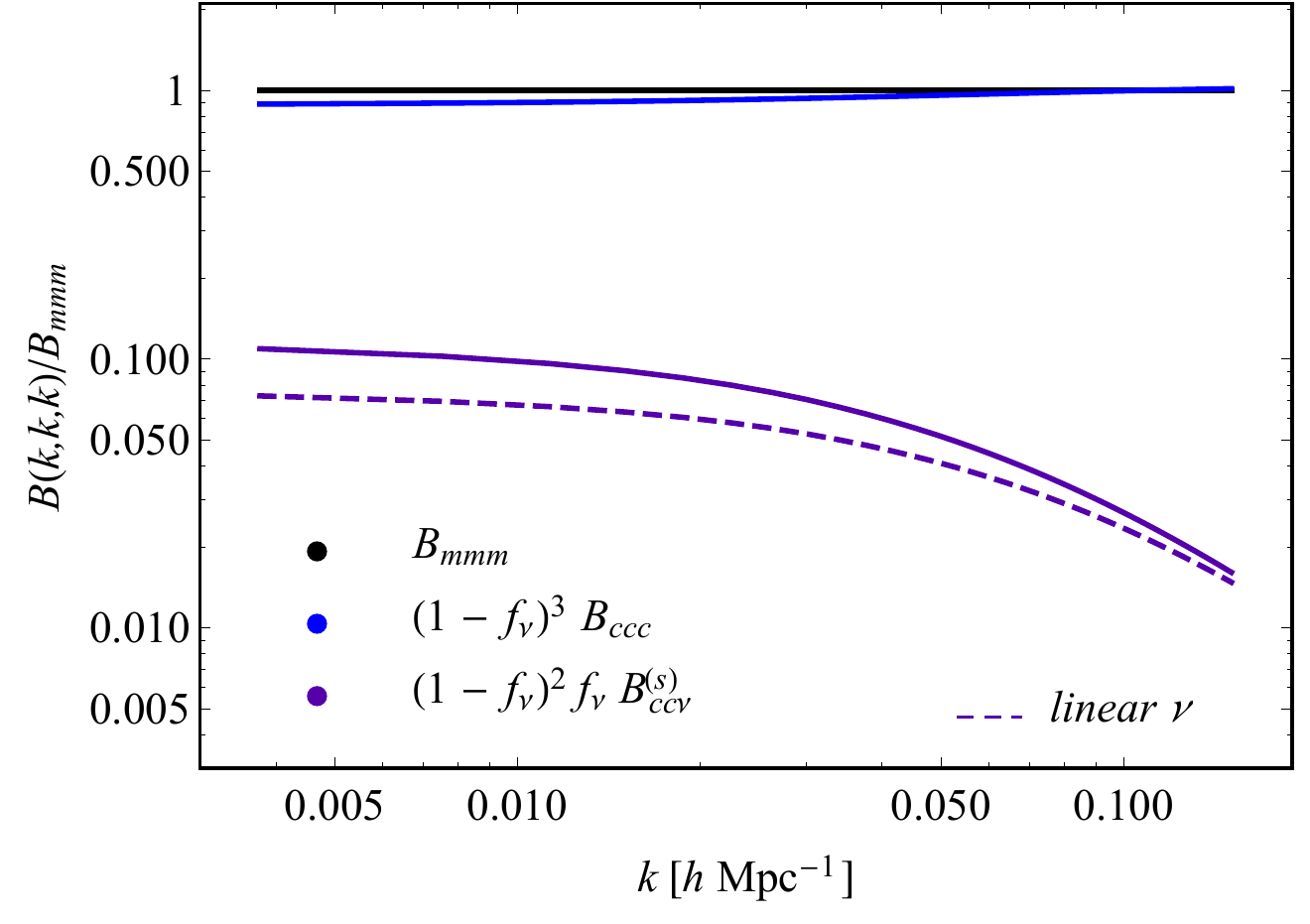}
\end{center} 
\caption{\label{fig:linear} Relative contributions to the equilateral configurations of the total matter bispectrum from the different terms in eq.~(\ref{eq:Bmmm}) up to $\mathcal{O}(f_\nu)$ for $m_\nu = 0.53$ eV at $z=0$. The dashed line shows the assumption neutrinos are linear on all scales. }   
\end{figure}

At tree-level in PT both the cold matter and the neutrino perturbations in eq.~\ref{eq:Bnnc} contribute to the neutrino-neutrino-cold matter component $B_{\nu\nu c}^{(s)}$, such that
\be
B^{(s),tree}_{\nu\nu c}(k_1,k_2,k_3) = B_{\nu\nu c, 112}^{(s)}(k_1,k_2,k_3)  +B_{\nu\nu c, 121}^{(s)}(k_1,k_2,k_3)  + B_{\nu\nu c, 211}^{(s)}(k_1,k_2,k_3) \,, 
\ee
where the subscripts ``112'' indicate the order of the perturbations $\d_\nu$, $\d_\nu$ and $\d_c$, respectively. It is easy to see that 
\begin{align}
\label{eq:PTBnnc}
& B_{\nu\nu c, 211}^{(s),}(k_1,k_2,k_3) = 2 \,F_2(\kv_1,\kv_2) P_{\nu c}^L(k_1) P_{\nu \nu}^L(k_2) + \rm{2~perm}\;\, \\
\label{eq:PTBnnc2}
& B_{\nu\nu c, 112}^{(s),}(k_1,k_2,k_3) =2 \,F_2(\kv_1,\kv_2) P_{\nu c}^L(k_1) P_{\nu c}^L(k_2) + \rm{2~perm}\;\,
\end{align}
and $B_{\nu \nu c , 121}^{(s)} =  B_{\nu\nu c, 211}^{(s)}$ (we are assuming the ordering of the subscript to correspond to the perturbations $\d_c$, $\d_c$ and $\d_\nu$ in this order). Setting neutrino perturbations to their linear value on {\em all scales} leads to $B_{\nu \nu c , 121}^{(s)} =  B_{\nu\nu c, 211}^{(s)} = 0$, resulting, in turn, in a $\mathcal{O}(1)$ bias in $B_{\nu\nu c}$ at scales $k<k_{\rm{fs}}$. At the one-loop level, the only contribution is coming from the fourth-order correction to $\d_c$ so that we only have 
\be
B_{\nu\nu c, 114}^{(s)}(k_1,k_2,k_3) = 4\,P_{\nu c}^L(k_1)\,P_{\nu c}^L(k_2)\, \int d^3q\,F_4(\qv,-\qv,\kv_1,\kv_2)\,P_{cc}^L+{\rm 2~perm.}\,.
\ee
The full prediction up to one-loop would then be given by
\be
B_{\nu\nu c}^{(s)}\simeq B_{\nu\nu c}^{(s),tree}+B_{\nu\nu c, 114}^{(s)}\,.
\ee
For the cold-cold-neutrino component $B_{cc\nu}^{(s)}$  the tree-level expression is the same of eq.~\ref{eq:PTBnnc} with the replacement $\nu\leftrightarrow c$ in all the terms.
At one-loop we have now three types of contributions,
\bea
B_{cc\nu, 411}^{(s)}  = B_{cc\nu, 141}^{(s)} & = & 4\,\left[P_{\nu c}^L(k_1)\,P_{c c}^L(k_2)+P_{\nu c}^L(k_2)\,P_{c c}^L(k_1)\right]\,\nonumber\\
& & \times \int d^3q\,F_4(\qv,-\qv,\kv_1,\kv_2)\,P_{cc}^L(q)+{\rm 2~perm.}\,,\\
B_{cc\nu, 321}^{(s)}  & = & 6\,P_{\nu c}^L(k_1) \int d^3q\,F_3(\kv_1,\kv_2-\qv,\qv)\,F_2(\kv_2-\qv,\qv)\,P_{cc}^L(|\kv_2-\qv|)\,P_{cc}^L(q)\nonumber\\
& & +{\rm 2~perm.}\,
\eea
with the additional $B_{cc\nu, 231}^{(s)}$ term obtained by exchanging $\kv_1$ with $\kv_2$, and
\be
B_{cc\nu, 312}^{(s)}   =  6\,P_{c c}^L(k_1) \int d^3q\,F_3(\kv_1,\kv_2-\qv,\qv)\,F_2(\kv_2-\qv,\qv)\,P_{c\nu}^L(|\kv_2-\qv|)\,P_{cc}^L(q)+{\rm 2~perm.}\,
\ee
with an analogous $B_{cc\nu, 132}^{(s)}$.
Up to one-loop correction we have then 
\be
B_{cc\nu}^{(s)}\simeq B_{cc \nu}^{(s),tree} + 2 B_{cc\nu, 411}^{(s)}+B_{cc \nu, 321}^{(s)}+B_{cc \nu, 231}^{(s)}+B_{cc \nu, 312}^{(s)}+B_{cc \nu, 132}^{(s)}\,.
\ee
The $B_{ccc}$ is then given by the usual PT expressions in terms of the cold matter linear power spectrum $P_{cc}^L$ \citep{BernardeauEtal2002}. 
We notice right away that one-loop corrections to the mixed contributions $B_{\nu\nu c}^{(s)}$ and $B_{cc\nu}^{(s)}$, as we will see in section~\ref{sec:bispm}, are essentially irrelevant. More important, instead, are the implications of the linear neutrinos approximation on the tree-level prediction, since the relevant short-comings will take place at large scales.
We can also anticipate that assuming linear neutrino perturbations will result in an error of order  $f_\nu$ on scales larger than the free-streaming scale, which could exceed the \% level for realistic value of neutrino masses. This is shown in Fig.~(\ref{fig:linear}) for $m_\nu = 0.53$ eV at $z=0$, where we plot, within tree-level PT, the relative contribution of terms up to $\mathcal{O}(f_\nu)$ to the total matter bispectrum of equilateral configurations. The assumption of linear neutrinos on all scales indeed yields an inconsistent result at low $k$, $B_{cc\nu}$ is underestimated by more than a factor of 2, with biases of several \%s on $B_{mmm}$. 

\section{The DEMNUni simulations suite}
\label{sec:nbody}

We make use of  the ``Dark Energy and Massive Neutrino Universe'' (DEMNUni) suite of N-body simulations \cite{CarbonePetkovaDolag2016}, 
representing one of the best set of simulations of massive neutrino cosmologies both in terms of mass resolution and volume.
A complete description can be found also in \cite{CastorinaEtal2015}, here we briefly summarize the main details. 

All the simulations assume a baseline cosmology according to the
\planck\ results~\cite{PLANCK2013parameters}, namely a flat \lcdm\ model
with $ h = 0.67 $ as Hubble parameter, $n_s = 0.96$  as primordial spectral index, and  
$A_s = 2.1265 \times 10^{-9}$ for the amplitude of initial scalar perturbations. 
This implies that simulations with massive neutrinos have lower value of $\sigma_8$ with respect to the $\Lambda$CDM case. 
The total matter energy density and the baryonic energy density are set to $\Omega_m = 0.32$ and $\Omega_b = 0.05$ for all cosmologies, 
while the relative energy densities of cold dark matter, $\Omega_c$ (and neutrinos, $\Omega_\nu$) 
vary for each model as $\Omega_c =0.27$,  $0.2659$,  $0.2628$ and  $0.2573$, for $m_\nu = 0$, $0.17$, $0.3$ and $0.53$ eV, respectively.

The DEMNUni simulations have been performed using the tree particle 
mesh-smoothed particle hydrodynamics (TreePM-SPH) code
\gadgetthree~\cite{SpringelEtal2005}, specifically modified by~\cite{VielHaehneltSpringel2010} to   
account for the presence of massive neutrinos. 
They are characterised by a softening length
$\varepsilon=20 \kpc$, start at $z_{in}=99$, and have being performed in a cubic box of side $L = 2000 \Mpc$, 
containing  $N_p = 2048^3$ CDM particles, and an equal number of neutrino particles when $m_\nu \neq 0 $.
These features make the DEMNUni set suitable for the analysis of different cosmological probes, 
from galaxy-clustering, to weak-lensing, to CMB secondary anisotropies. 
  
Halos and sub-halo catalogs have been produced for each of the 62 simulation particle snapshots, via the friends-of-friends (FoF) and SUBFIND algorithms
included in \gadgetthree~\cite{SpringelYoshidaWhite2001,DolagEtal2009}. The linking length was set to be $1/5$ of the mean inter-particle
distance \citep{DavisEtal1985} and the minimum number of particles to identify a parent halo was set to 32,
thus fixing the minimum halo mass to $M_{\rm FoF}\simeq2.5\times 10^{12}$\hmone $M_\odot$.
In this work we consider three lower threshold for the halo mass, $M > 10^{13}$\hmone $ M_\odot$ , $M >3\times 10^{13}$\hmone $ M_\odot$  
and $M > 10^{14}$\hmone $ M_\odot$ and three values for the snapshot redshift,   $z = 0$,  $z = 0.5$, $z = 1$.

\section{The matter bispectrum}
\label{sec:bispm}

We measure the total matter bispectrum $B_{mmm}$ along with all its individual components $B_{ccc}$, $B_{cc\nu}^{(s)}$, $B_{\nu\nu c}^{(s)}$ and $B_{\nu\nu\nu}$ as defined by eq.~(\ref{eq:Bmmm}). For all components we consider all triangular configurations defined by discrete wavenumbers multiples of $\Delta k=3 k_f$ with $k_f\equiv 2\pi/L$ being the fundamental frequency of the box, up to a maximum value of $0.38\kMpc$.  The estimator of the bispectrum is given by
\be
\hat{B}(k_1,k_2,k_2)\equiv\frac{k_f^3}{V_B(k_1,k_2,k_3)}\int_{k_1}\!\! d^3q_1\int_{k_2} \!\!d^3q_2\int_{k_3} \!\!d^3q_3\,\d_D(\qv_{123})\,\d_{\qv_1}\,\d_{\qv_2}\,\d_{\qv_3}
\ee
where the integrations are taken on shells of size $\Delta k$ centered on $k_i$ and where
\be
V_B(k_1,k_2,k_2)\equiv \int_{k_1}\!\! d^3q_1\int_{k_2} \!\!d^3q_2\int_{k_3} \!\!d^3q_3\,\d_D(\qv_{123}) \simeq   8\pi^2\ k_1 k_2 k_3 \Delta k^3
\ee
is a normalisation factor counting the number of fundamental triangles in a given triangle bin. Its implementation is based on the algorithm described in \citep{Scoccimarro2015} and taking advantage of the aliasing reduction technique of \citep{SefusattiEtal2016}. As only one realisation for each cosmology is available, all error bars shown correspond to the Gaussian prediction given, for a generic bispectrum, by \citep{Scoccimarro2000B}, 
\be
\Delta B^2 (k_1, k_2, k_3)  \simeq s_B \frac{k^3_f}{V_B }  P(k_1)P(k_2) P(k_3)\,,
\label{eqerror}
\ee
with $s_B = 6, 2, 1$ for equilateral, isosceles and scalene triangles respectively.

\begin{figure}[t!]
\begin{center}
\includegraphics[width=.9\textwidth]{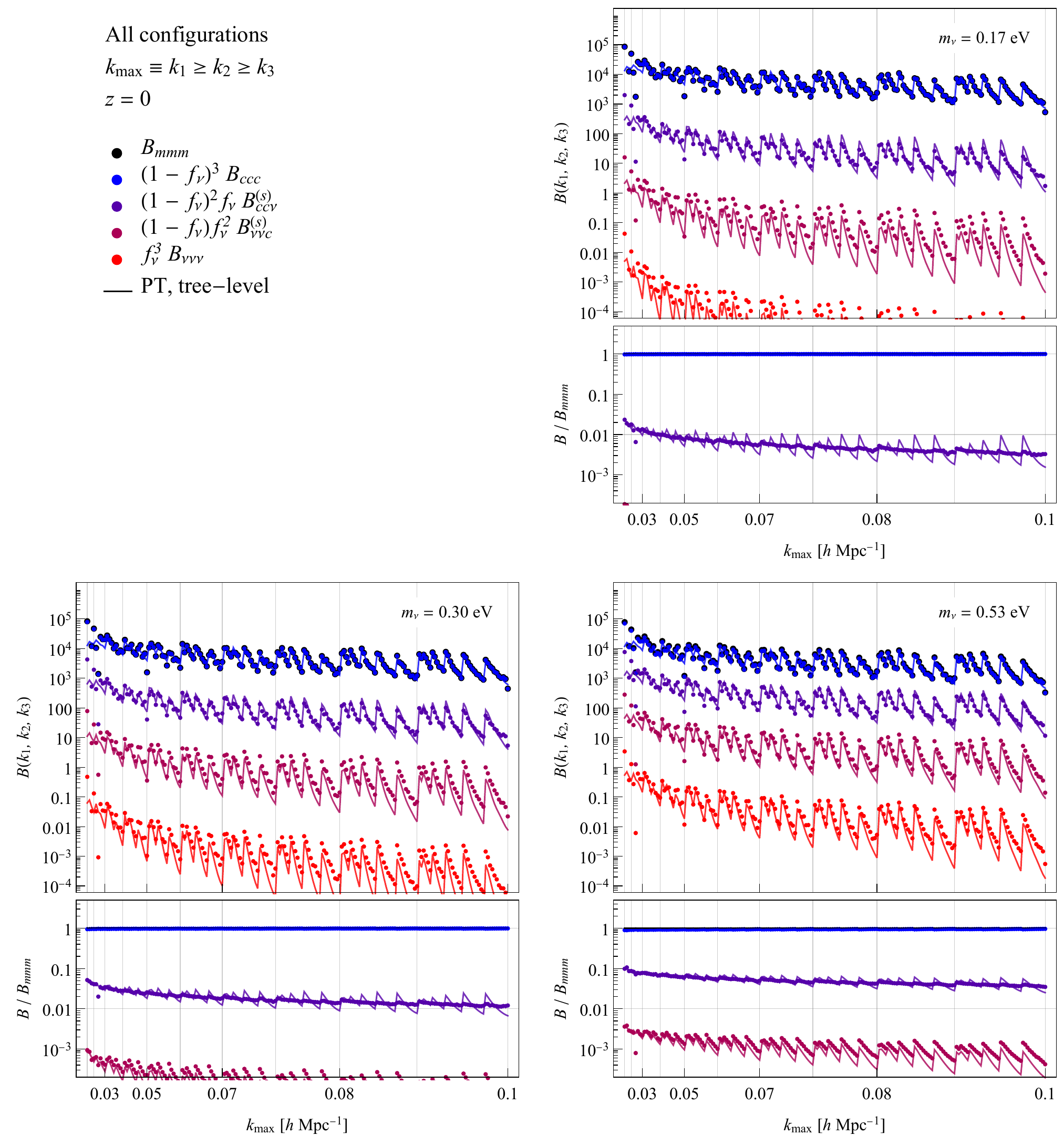}
\end{center} 
\caption{\label{fig:componentsall} Measurements of all the components of the total matter bispectrum $B_{mmm}(k_1,k_2,k_3)$ at $z=0$, properly weighted, compared with the tree-level prediction in PT. All triangular configurations re ordered with increasing $k_1$ and assuming $k_1\ge k_2\ge k_3$. In each panel, the bottom-half shows the relative contribution of each component to $B_{mmm}$ (only the relevant ones, contributing above the $0.1\%$ level are shown). Vertical lines correspond to equilateral configurations. }   
\end{figure}

Figure \ref{fig:componentsall} shows the  measurements at $z=0$ of all configurations for all components, {\em rescaled} by the proper factors as a function of the neutrino fraction, according to eq.~(\ref{eq:Bmmm}), in order to  assess directly the relative contribution to the total matter bispectrum. The bottom half of each panel shows the ratio of each component to the total $B_{mmm}$. Triangles are ordered with increasing $k_1$ and assuming $k_1\ge k_2\ge k_3$ so that all configurations shown correspond to large-scales with $k_1\le 0.1\kMpc$. Data points from N-body simulations are compared to tree-level predictions in PT. Theoretical predictions are computed for ``effective''  values of the wavenumbers defined, for a given configuration of sides $k_1$, $k_2$ and $k_3$  by
\be
\tilde{k}_{1,23}\equiv \frac1{V_B}\int_{k_1}\!\! d^3q_1\,q_1\,\int_{k_2} \!\!d^3q_2\int_{k_3} \!\!d^3q_3\,\d_D(\qv_{123})\,,\ee
and similarly for the other two values. Differences with respect to evaluations at the center of each $k$-bin are marginally relevant only for the largest scales.

The first, rather obvious, observation is that the only relevant, \ie above 1\% level, contribution to the total matter bispectrum $B_{mmm}$, in addition to the cold matter bispectrum $B_{ccc}$, is given by the cross-bispectrum $B_{cc\nu}^{(s)}$ which is of $\mathcal{O}(f_\nu)$. Therefore a proper theoretical description of the matter bispectrum in massive neutrinos cosmologies should focus on both these two components. Here we show a first comparison with tree-level PT, providing rather accurate predictions to the cold matter component $B_{ccc}$ over the scales shown in the figure. Predictions for the $B_{cc\nu}^{(s)}$ and $B_{\nu\nu c}^{(s)}$ are very well described by the assumption neutrino contribute only at tree-level in PT.

As we consider smaller scales, in the quasi-linear regime, the tree-level approximation becomes more accurate for the cross-bispectra of cold matter and neutrinos because of small-scale suppression of neutrino perturbations. On the other hand, we expect further nonlinear (one-loop) corrections to become important only for the cold matter contribution $B_{ccc}$.  This is the natural extention to higher-order correlation functions of previous results for the total matter power spectrum, where linear theory was sufficiently accurate to describe the neutrinos and cross neutrino-cold matter components $P_{\nu\nu}$ and  $P_{c\nu}$, respectively \citep{SaitoTakadaTaruya2008, BlasEtal2014, CastorinaEtal2015}. 

\begin{figure}[t!]
\begin{center}
\includegraphics[width=0.9\textwidth]{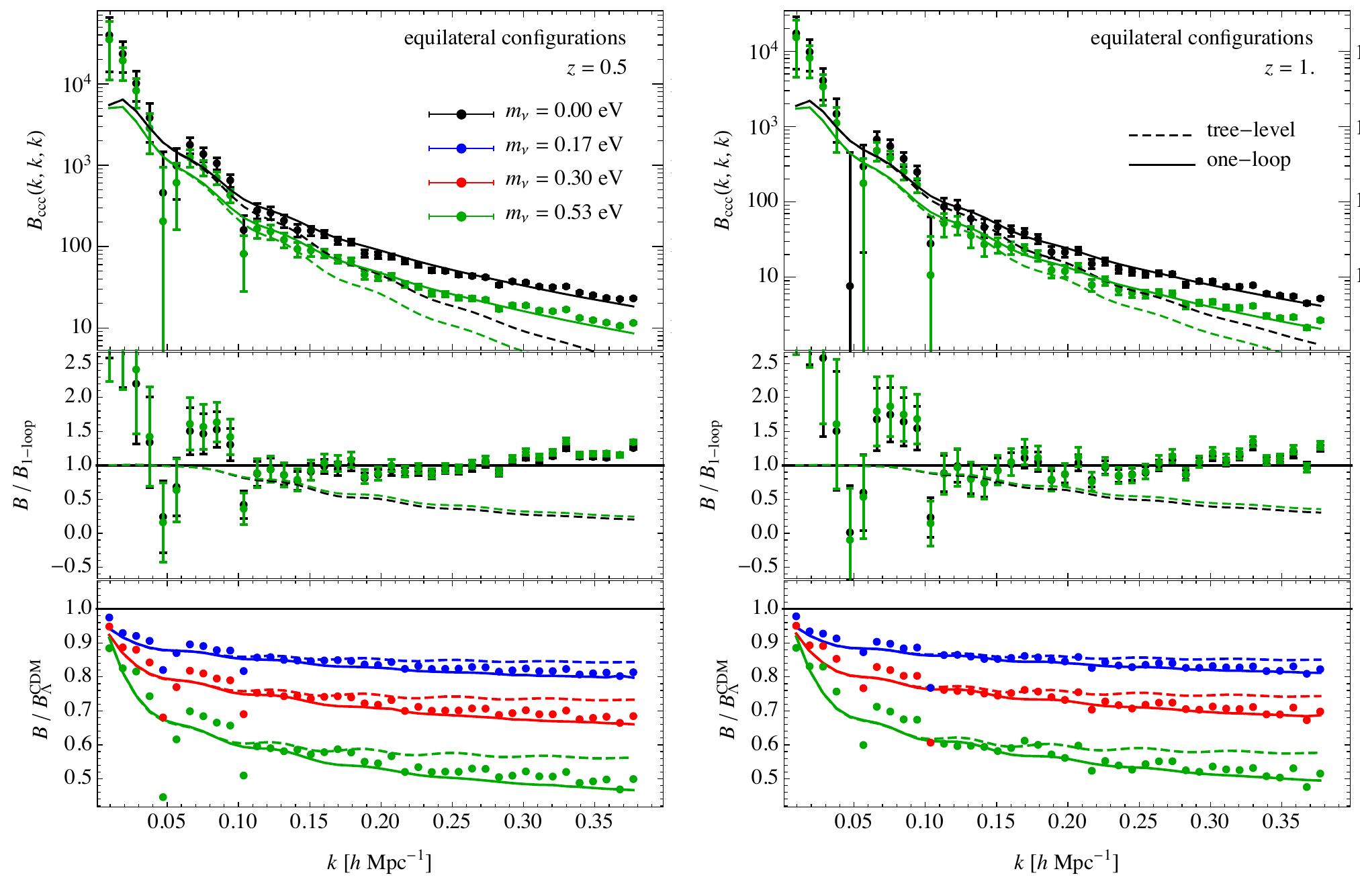}
\includegraphics[width=0.9\textwidth]{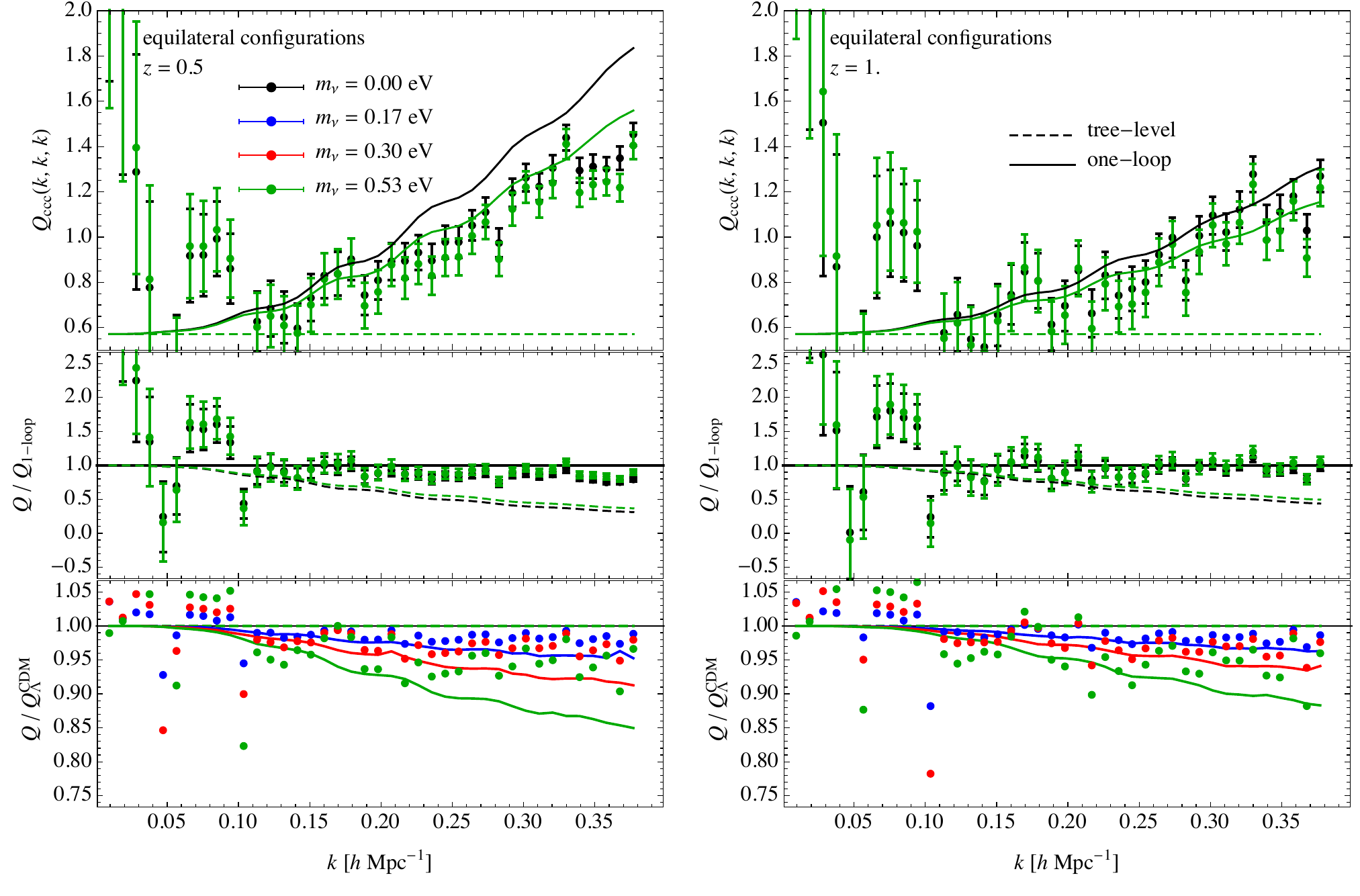}
\end{center} 
\caption{\label{fig:BcEq} Top panels: equilateral configurations of the cold matter bispectrum, $B_{ccc}(k,k,k)$, compared with tree-level (dashed curves) and one-loop (continuous curves) predictions in PT. Bottom panels: same comparison for the {\em reduced} cold matter bispectrum $Q_{ccc}(k,k,k)$. Left panels show the results at $z=0.5$, right panels at $z=1$. In addition to the measurements of $B$ or $Q$, we plot  the residuals with respect to the one-loop predictions (middle panel) and the ratio between the $m_\nu\ne0$ cosmologies to the corresponding massless neutrino case (bottom panel). } 
\end{figure}

Figure~\ref{fig:BcEq} shows the measurements of the cold matter bispectrum, $B_{ccc}$, at $z = 0.5$ (left panel) and $z=1$ (right panel), for equilateral triangles (left panel). We compare with analytical predictions at  tree (dashed) and 1-loop (continuous) level in PT \citep{BernardeauEtal2002,Sefusatti2009}. Different colors indicate different value of the sum of neutrino masses , $m_\nu =  0,\,0.3,\, 0.53\,\text{eV}$  (black, red and green respectively). 

The analytic curves have been obtained according to the prescription as in \citep{CastorinaEtal2015}; we consider the perturbative kernels in cosmology with massive neutrinos to have the same form as in  $\Lambda$CDM cosmology and we assume all the effects induced by neutrinos encoded in the linear power spectrum. Our assumptions are justified by previous studies \citep{SaitoTakadaTaruya2009, CastorinaEtal2015} which showed this approximation to work better than a $ \%$ on the power spectrum analysis. 
Middle panels show the ratio between the data for the three different cosmologies ($\Lambda$CDM  and $m_\nu = 0.53$ eV) with respect to their 1-loop predictions (black and green points); 
we also plot the ratio between tree-level and 1-loop prediction with dashed lines. The approximation of neglecting the effects of massive neutrinos on the 1-loop CDM bispectrum other than the different linear theory $P_{cc}(k)$, provides the same level of agreement we find in the $\Lambda$CDM case.
In the bottom panels we present the ratio between the bispectrum measured for $m_\nu \ne  0$ with respect to the  $\Lambda$CDM measurement;  continuous and dashed lines denote instead the same ratio as predicted at the 1-loop and tree-level in PT. The comparison with the power spectrum analysis reveals, as expected, that equilateral configurations are roughly two times more sensitive than the power spectrum to massive neutrinos. We also notice that the suppression of CDM bispectrum with respect to the standard case does not evolve significantly with redshift, in agreement with well known results at the two-point function level. As expected \citep{ScoccimarroEtal1998, SefusattiCrocceDesjacques2010} one-loop predictions in standard PT tend to overestimate the measurements in equilateral configurations at low redshift. We find a good agreement up to $k=0.3\kMpc$ at $z=0.5$, but it should be kept in mind that the precision of our measurement is roughly 10\% at these scales. 

A better agreement with measurements can clearly be found in the context of the Effective Field Theory of the Large-Scale Structure \citep{CarrascoHertzbergSenatore2012}, see \citep{AnguloEtal2015B} for its application to the matter bispectrum. 
We limit ourselves to compare the accuracy of standard PT predictions in massive neutrino cosmology with known results for the $\Lambda$CDM case. 

\begin{figure}[t!]
\begin{center}
\includegraphics[width=0.9\textwidth]{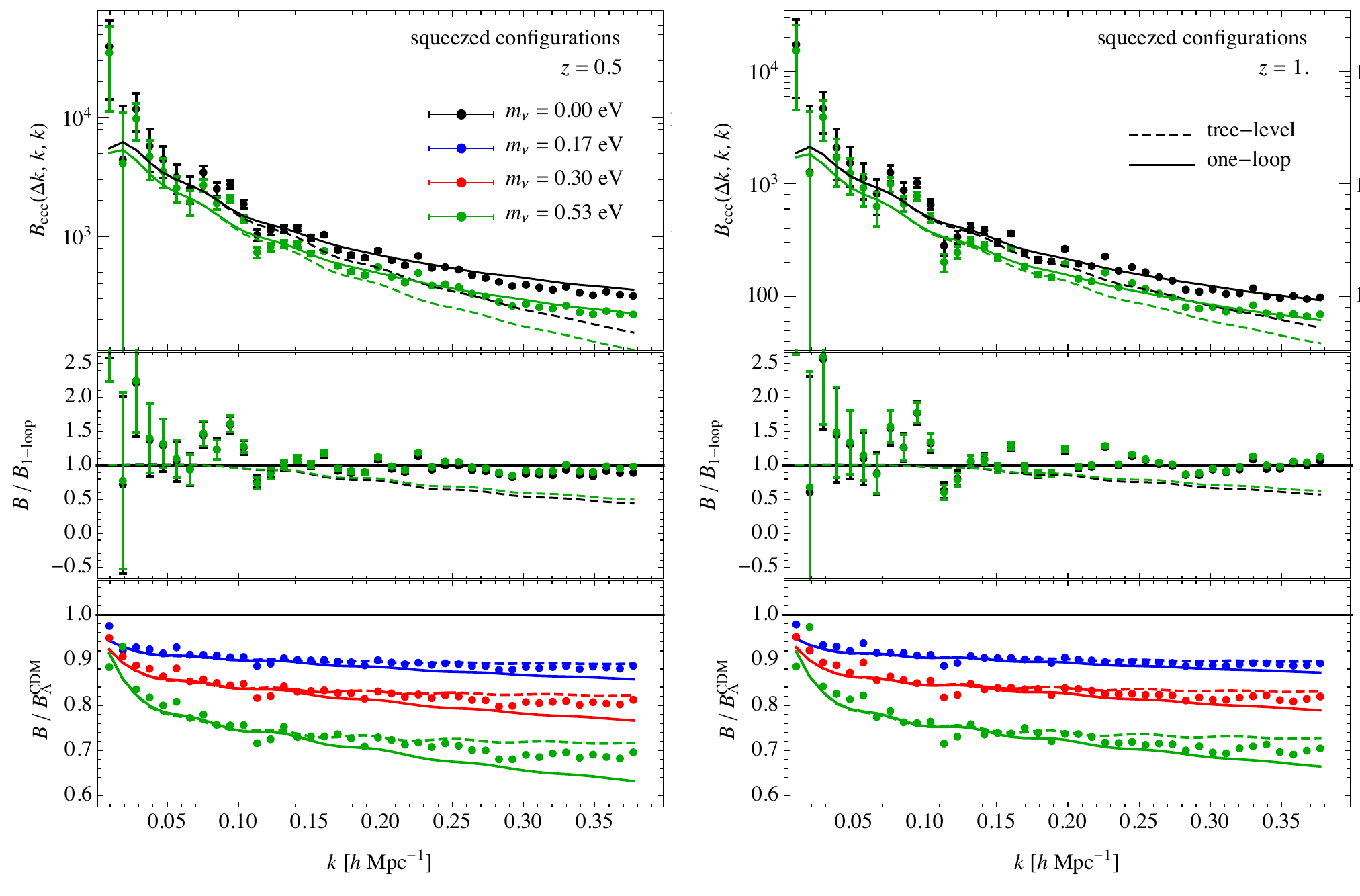}
\includegraphics[width=0.9\textwidth]{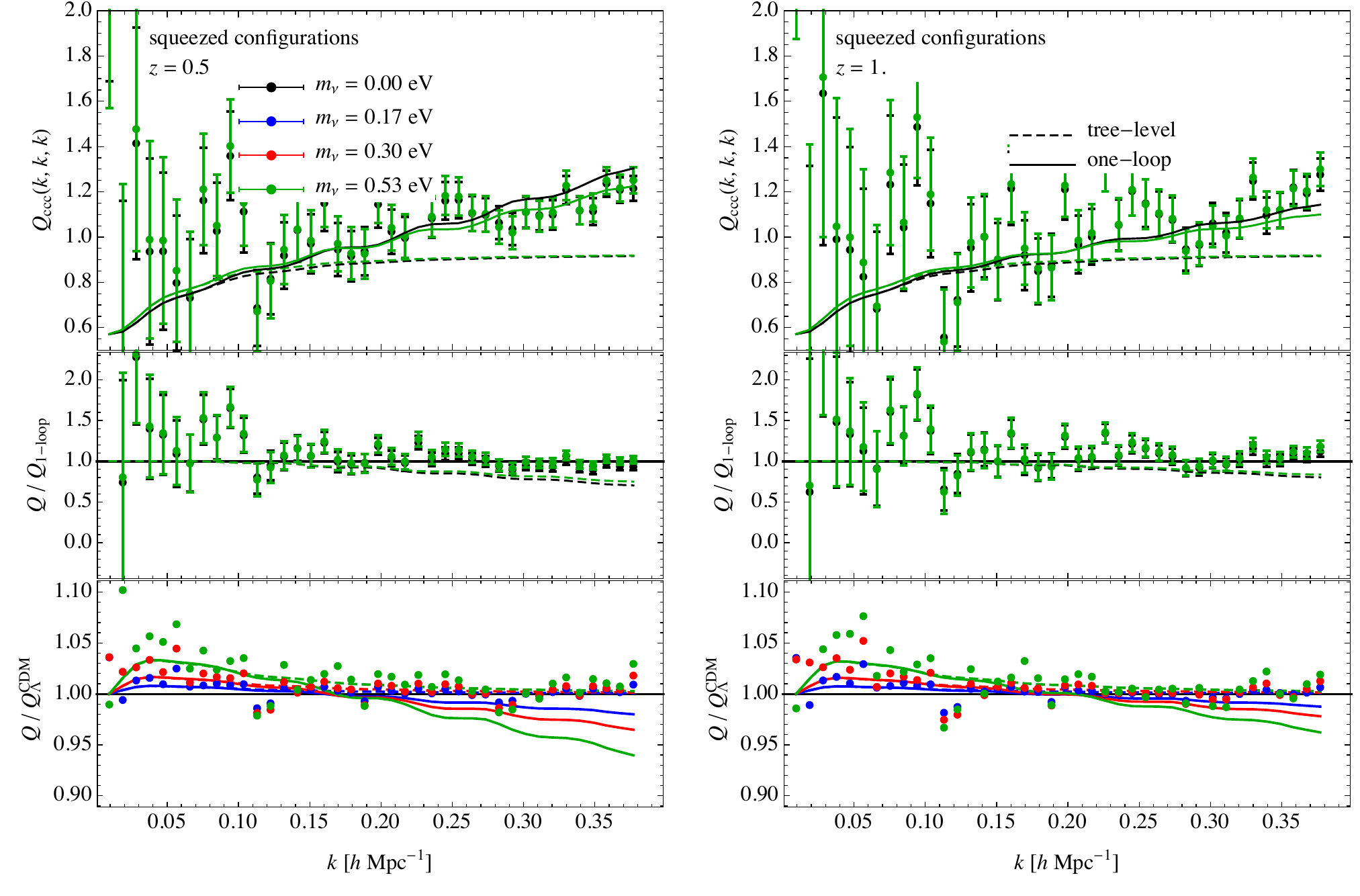}
\end{center} 
\caption{\label{fig:BcSq} Same as figure~\ref{fig:BcEq} but for squeezed configurations $B_{ccc}(k_l,k,k)$ with fixed $k_l=0.009\kMpc$ as a function of $k$.}
\end{figure}

More generally, the suppression of the amplitude of the bispectrum in cosmology with massive neutrinos compared to a standard cosmological model is a function of the triangle shape. 
For instance it is easy to see that squeezed triangle configurations, $B_{mmm}(q, k , k )$  with $q\ll k$ are less affected by massive neutrinos. 
The phyisical interpretation of this is simple: for significantly large scales, $q$, the neutrinos behaves like CDM and therefore the relative suppression with respect to the $\Lambda CDM$ case is reduced compared to other triangular configurations.
At a scale where $q \ll k_{fs}$  then $\delta_{\nu}(q)\simeq \delta_{c}(q)$ and one has $ B_{\nu cc}(q , k , k ) \simeq  B_{c cc}(q , k , k )$ on all scales, even below free-streaming. 
While our perturbation theory calculations provide a reasonable fit to the measurements in the N-body, our analysis is in disagreement with the prediction of \citep{LeviVlah2016} who have analytically found that squeezed configurations show the largest neutrinos-induced suppression. Squeezed configurations are shown in Figure~\ref{fig:BcSq} with the same notations as Figure~\ref{fig:BcEq}. The overall accuracy of PT predictions is similar to the equilateral case and we can notice, again, a significant improvement of one-loop predictions over tree-level ones.

\begin{figure}[t!]
\begin{center}
\includegraphics[width=0.9\textwidth]{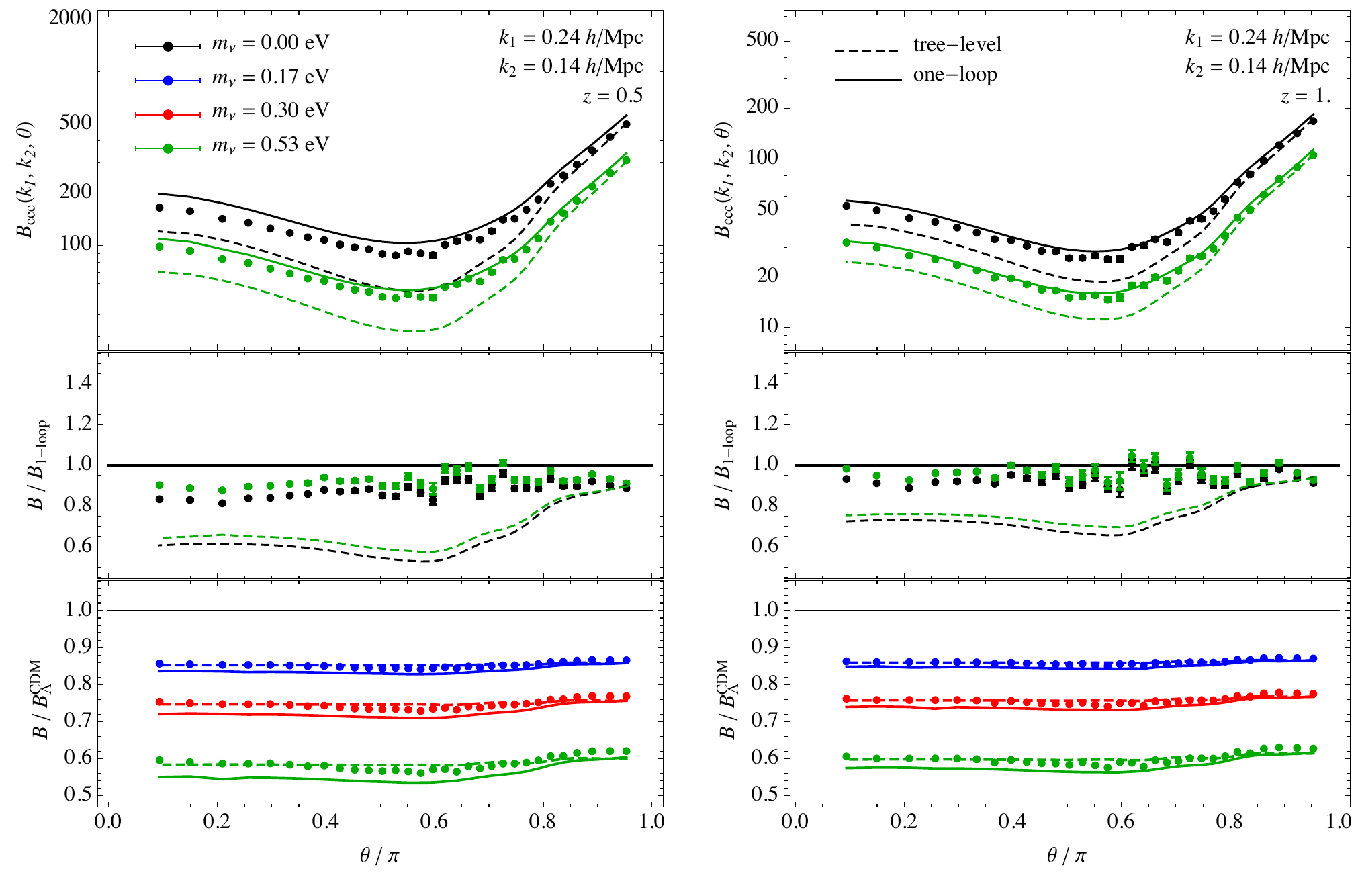}
\includegraphics[width=0.9\textwidth]{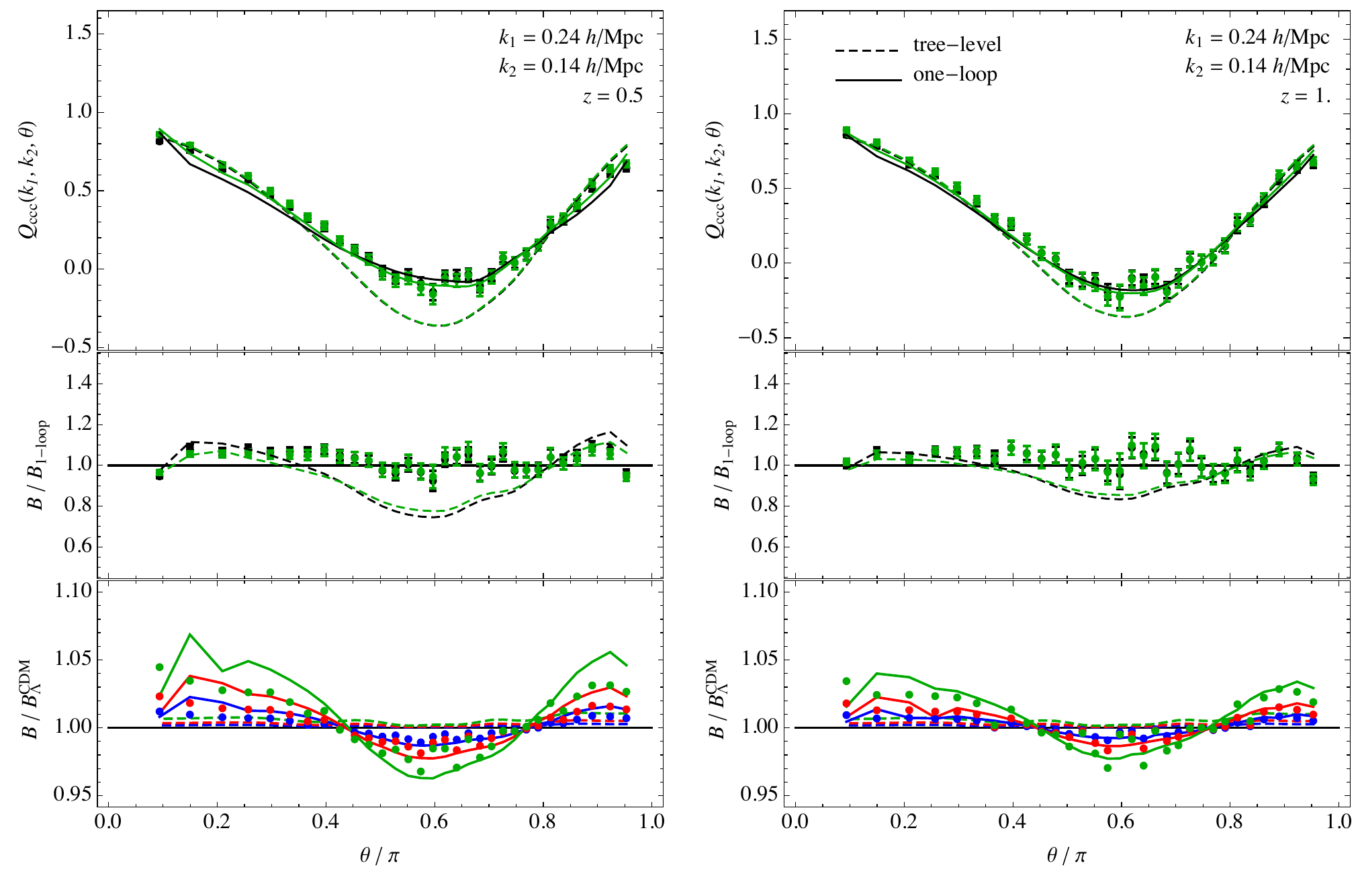}
\end{center} 
\caption{\label{fig:BcSc} Same as figure~\ref{fig:BcEq} but for scalene configurations with fixes sides $k_1=0.14\kMpc$ and $k_2=0.23\kMpc$ as a function of the angle between $\kv_1$ and $\kv_2$. }
\end{figure}

As another example of the effects massive neutrinos have on the CDM bispectrum, in Fig. \ref{fig:BcSc} we present the bispectrum for scalene triangles as a function of the angle $\theta$ between the two sides for all the cosmologies at $z =0.5$ (left panel) and redshift $z= 1.0$ (right panel).  We select triangle configurations in the mildly nonlinear regime, with $k_1 = 0.14\;h\;Mpc^{-1}, k_2 = 0.23\;h\;Mpc^{-1} $. The labels, colors and plots order correspond to those in figure \ref{fig:BcEq}.  Differently from the previous cases, the bispectrum in massive neutrino cosmologies is suppressed on all scales shown. PT is able to reproduce the measurements in the different cosmologies quite well at $z=1$ and large values of $\theta$ (the ``squeezed limit'') while it shows the usual overestimation, about 10-15\%, for more equilateral triangles and lower redshift.

An additional way to assess the effect of massive neutrinos on the bispectrum and compare it to the effect on the power spectrum, is to compute the reduced bispectrum $Q$ defined as
\be
Q(k_1,k_2, k_3) =  \dfrac{B(k_1, k_2, k_3)}{ P (k_1)P (k_2) + P (k_1)P (k_3) + P (k_2)P (k_3)}. 
\ee
This quantity, at tree-level in PT, does not depend on the initial amplitude of the linear fluctuations (or $\sigma_8$ in a $\Lambda$CDM cosmology), including the suppression due to neutrinos in linear theory, and it highlights the different level of nonlinearity in the bispectrum w.r.t. the power spectrum.
In all the figures~\ref{fig:BcEq}, \ref{fig:BcSq}  and \ref{fig:BcSc} the bottom part shows the reduced bispectrum for the choosen configurations at $z = 0.5$ and 1. Again, top panels show measurements for $Q$ compared with  tree-level and one-loop predictions; middle panels show the residuals with respect to  one-loop, while in the bottom panels we display the ratios between the reduced bispectrum measured for $m_\nu \ne  0$ with respect to $\Lambda$CDM. The errors are computed by propagating  $\Delta B$ and $\Delta P$ errors neglecting the cross-correlation between $P$ and $B$. 

The  small deviations we see in the bottom panels among different cosmologies indicate that any new, nonlinear neutrinos signature specific for the bispectrum is significantly small at scales $k>0.2\kMpc$. This implies that the bispectrum alone is not able to probe new physical effects induced by massive neutrinos in the clustering of dark matter; however it still represents a relevant asset as it breaks part of the degeneracy between the cosmological parameters, when combined with the power spectrum.
 One-loop predictions for $Q$  are well within the precision of our measurements. They are also qualitatively consistent with results for the relative effect of neutrino masses shown in the bottom panels, with the exception of the squeezed configurations at small scales where they overpredict significantly a neutrino signature not detectable in the measurements. These are, however, few percent effects, too small to be properly investigated with the limited statistic of a single set of simulations.

\section{The halo bispectrum}
\label{sec:bisph}

\subsection{Bias modeling}

According to the Eulerian bias model, at large scales,  the halo density field $\delta_h$ can be locally described as a function of the underlying smoothed density contrast $\delta$,  \citep{FryGaztanaga1993, BernardeauEtal2002, DesjacquesJeongSchmidt2016}. 
In particular if $\delta \ll 1$, we can model  $\delta_h$ as a taylor expansion in $\delta$,
\begin{equation}
\delta_h = \sum_{n} \frac{b_n}{n!}\delta^n,
\label{eulocbias}
\end{equation}
where the $b_n$ correspond to the bias parameters.
In this framework, the halo power spectrum, $P_{hh}$ or the halo-matter cross correlation, at very large scales, are related to the matter power spectrum $P(k)$, through 
\be
P_{hh}(k) = b_1^2 P(k)\,\quad P_{hm} = b_1 P(k)
\label{eq:halop}
\ee 
In a local Eulerian bias model the tree-level halo bispectrum reads
\be
B_{hhh} (k_1, k_2, k_3) =   b_1^3 B (k_1, k_2, k_3) + b_2 b_1^2 \Sigma_{123} (k_1, k_2, k_3) ,
\label{halobisp}
\ee
with $B$ being the matter bispectrum, $\Sigma_{123} \equiv P(k_1)P(k_2) + 2~\rm cyc$ and $b_2$  a quadratic bias parameter. The equation above shows that a measurements of the halo bispectrum on large scale could be used not only to constrain cosmological parameters, but also to break the degeneracy between the bias parameters and the amplitude of fluctuations in a power spectrum analysis.

It is well known,  see {\eg} \citep{Scoccimarro2000B, PollackSmithPorciani2012},  that fitting  $B_{hhh}$ with model in eq.~\ref{halobisp} yields different values of $b_1$ with respect to ones obtained from the halo power spectrum, modeled as in eq.~\ref{eq:halop}. Recent works on bias modelling \citep{BaldaufEtal2012, ChanScoccimarroSheth2012, ShethChanScoccimarro2013}  have shown the intrinsic mistake in considering the bias to be deterministic and local: the nonlinear evolution induced by gravity introduces new, non-local bias contributions proportional to operators built from derivatives of the density field and/or the gravitational potential. In this framework the halo density, at second order in the bias expansion,  takes the form
\be
\label{eq:brel}
\delta_h = b_1 \delta + \frac{b_2}{2} \delta^2 + \gamma_2\,\mathcal{G}_2, 
\ee
where $\mathcal{G}_2  $ is defined as
\be
\mathcal{G}_2 \equiv (\nabla_{ij} \Phi_v)^2 -  (\nabla_{ij}\Phi_v)^2\,, 
\ee
with $\Phi_v$ being the velocity potential such that ${\bf v} = \nabla \Phi_v$. 
The resulting tree-level halo bispectrum reads 
\be
B_{hhh} (k_1, k_2, k_3) =   b_1^3 B(k_1, k_2, k_3) + b_2 b_1^2 \Sigma_{123}(k_1, k_2, k_3)   + 2 \gamma_2 b_1^2 K_{123}(k_1, k_2, k_3) \,,
\label{halobispnonloc}
\ee
where $K_{123} \equiv (\mu_{12} -1) P(k_1)P(k_2) + 2\, \rm cyc $, $\mu_{12}$ the cosine of the angle between $\kv_1$ and $\kv_2$.

This model has been widely tested in $\Lambda$CDM cosmologies, and in this section we would like to extend these results to a cosmological model that include massive neutrinos. As first noted in \citep{CastorinaEtal2014}, the linear bias in a cosmology with massive neutrinos is scale independent only if the CDM power spectrum appears on the right hand side of \ref{eq:halop}. This is a consequence of the fact that neutrinos do not cluster on halos or galaxies scales, and therefore fluctuations in the number of halos and galaxies respond to the CDM field only. 

Given this result, we will model the halo bispectrum and then fit for the bias parameters in the next section assuming the halo bias relation in \ref{eq:brel} is written in terms of the CDM field only.  Recovering the same linear bias of the power spectrum from the bispectrum would yield a confirmation of the correctness of the argument for linear bias in \citep{CastorinaEtal2014}.

\subsection{Fitting procedure}

\begin{figure}[t]
\begin{center}
\includegraphics[width=.9\textwidth]{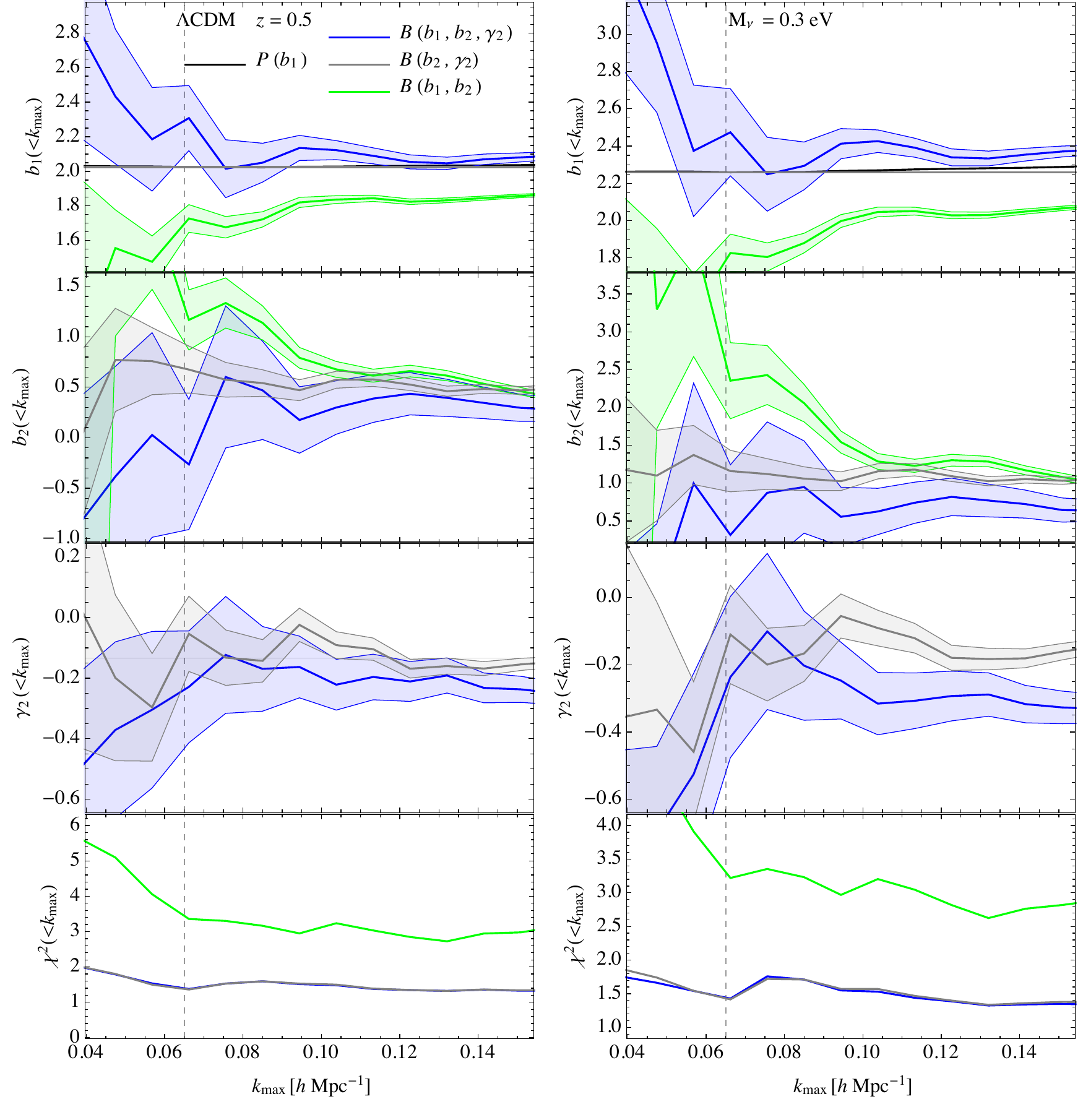}
\end{center} 
\caption{ \label{fi2} 
Tthe best fits bias parameters in $\Lambda$CDM framework (left panel) and neutrino cosmologies $m_\nu = 0.53 \;eV $,  (right panel), assuming a local  and  non local bias model, at $z= 0.5$. 
Green lines indicate the measurements of  $b_1$  $b_2$  for the local model, (top and middle-top panels). Blue color indicates  $b_1$,  the \textit{effective} $b_2$ and the non local corretion $\gamma_2$ (middle-bottom panels) to the bias model. Black lines correspond  $b_1$ as measured from the halo power spectrum $P_{hh}$. 
Grey color shows the best fits for $b_2$ and $\gamma_2 $  fixing  $b_1$  from the power spectrum fits.
In the bottom panels we show the $\chi^2$ related to the different fitting formulas. 
}   
\end{figure}

We measure the halo bispectrum in a $\Lambda$CDM cosmology and in cosmologies with all the three different values of neutrino masses, $m_\nu = 0.17, 0.3, 0.53$ eV. We do not consider a part of the measurements for the high-mass threshold,  $M > 10^{14} M_\odot/h$, and redshift $z = 1.0 $ since they are highly affected by shot-noise contributions, which for the halo bispectrum take the following form
\be
\hat{B}_{hhh}(k_1,k_2,k_3) = B_{hhh}(k_1,k_2,k_3) + \frac{1}{\bar{n}}[P_{hh}(k_1) + 2 \,\text{cyc}] + \frac{1}{\bar{n}^2} 
\ee

We compare the best fit measurements of $b_1$ and $b_2$ assuming a local model for the bias  the values for  $b_1$, $b_2$ and $\gamma_2$ when assuming the non local model.  At the level of  two-point statistics we fit a linear model to the halo-matter cross-correlation, $P_{hm}(k)$, as a function of the maximum wavenumber $k_{max}$.  For the bispectrum we compute the likelihood $\mathcal{L} $ up to a value $k_{max}$ for the largest side for each triangle configuration. In the case of the non local model we have, 
\be 
\ln \mathcal{L} =\sum_{k }^{ k_{max}} \frac{   B_{hhh} - b_1^3 B_c - b_2 b_1^2 \Sigma_{123}  - 2 \gamma_2 b_1^2 K_{123} - \alpha_1(k_1^2 + k_2^2) (P_1 P_2) + 2\rm\;cyc}{\Delta B^{2}_{hhh} }\,,
 \ee
whith $\Delta B^{2}_{hhh}$ as Eq. (\ref{eqerror}). A similar expression is used to estimate the linear bias from the halo-matter cross power spectrum, for which we also assume Gaussian covariance. Possible, unaccounted scale-dependences in the modeling are marginalized over the amplitude $\alpha$ of a generic $k^2$-like term in both the likelihood for the power spectrum and the bispectrum. This term will take care of higher loop corrections and scale dependent bias on large scales \cite{EFT,Desjacques}. All the fits have been computed using the downhill simplex method, see {\eg} \citep{PressEtal1992}. We determined the errors on the best-fit values using the Fisher matrix prescription, see for instance \citep{TegmarkTaylorHeavens1997}.   
In order to compare the results for $b_2$  from local {\em vs} non local models,  we introduce an effective local quadratic bias, $b_{2, \rm eff} = b_2 - 4/3 \gamma_2$, \citep{ChanScoccimarro2012},  in case of the non local prescription, corresponding to the amplitude of the monopole component of the overall quadratic bias correction. 
For simplicity of notation, since we do not make use of the quadratic bias in the case of non local prescription,  we will refer to $b_{2, \rm eff}$ as $b_2$, omitting the second subscript. The CDM bispectrum in the above equation is computed at $1$-loop in perturbation theory, whereas the other terms are evaluated at tree-level, so we expect any failure of the theoretical predictions at low scales to be mainly due to the bias model. 

In figure~\ref{fi2} we present  the best fits of the bias parameters from the measurements in two different cosmologies, $\Lambda$CDM (left panels) and with $M_\nu = 0.3$ eV (right panels). Both figures are organised in the same way: the measurements of  $b_1$ and $b_2$  at $z= 0.0$ (left panel) and at $z= 0.5$ (right panel); cyan and orange colours indicate  $b_1$ and $b_2$ for the local model while blue and red colours indicate  $b_1$ and \textit{effective} $b_2$ for the non local prescription. Black lines correspond to the measurement of $b_1$ obtained from the halo-matter power spectrum $P_{hm}$. We noticed that the measurements of $b_1$ in a local bias framework disagree with the estimates from two-point statistics, but that on the other hand when we account for $\gamma_2$, we find a good accordance; this confirms the effect of non-local bias on the halo bispectrum measurements even in a massive neutrino cosmology. It also reinforces our understanding of halo/galaxy bias in these cosmologies in terms of hte CDM field only.

\begin{figure}[t]
\centering
\includegraphics[width=1. \textwidth]{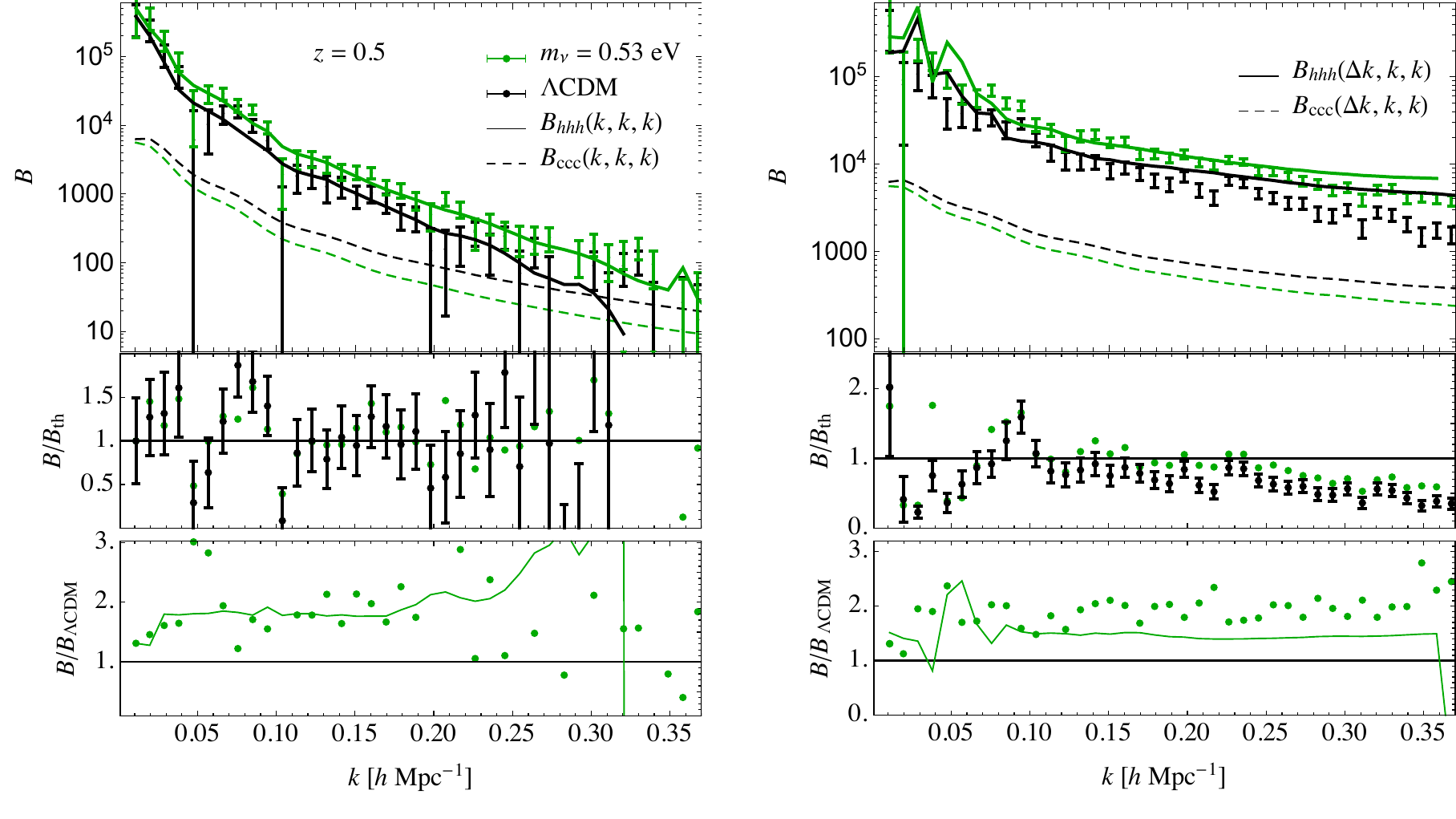}
\caption{\label{bispectrumhalo}  The measurements  of the Bispectrum for equilateral and squeezed configurations. Dashed lines show the theoretical predictions for matter bispectrum in  $\Lambda$CDM (black) and in cosmology with $M_\nu = 0.53eV$ (green).
Continuous lines indicate the predictions for the halo bispectrum assuming a non local bias model; black and green points mark the measurements of the halo bispectrum at  $M >10^{13} M_\odot$.}   
\end{figure}

Even though the error budget of our measurements of the Bispectrum in the simulations does not allow us to test the bias model of the Bispectrum at the \% level, it is still interesting to check down to which scale our bias model provides a reasonable fit to the bispectrum in the N-body. On the baryonic acoustic oscillations scales our errors are indeed comparable with those in the real data from \citep{GilMarinEtal2017}.  Figure~\ref{bispectrumhalo} shows the measurements  of the bispectrum for equilateral and squeezed configurations. We selected the halo mass range $A$ at redshifts $z = 0.5$. The dashed lines show the theoretical predictions for the matter bispectrum in case of $\Lambda$CDM (black) and cosmology with $M_\nu = 0.53 \;eV$ (green).
The continuous lines indicate the predictions for the halo bispectrum assuming a non local eulerian model for the bias. The agreement of the model with the data breaks down earlier in the presence of massive neutrinos, as the bias parameters are higher and therefore nonlinearities stronger. Nevertheless we find our bias template is able to fit the measured halo bispsctrum down to $k\simeq 0.15 \kMpc$. 
It is important also to notice that in comparison with thhe $\Lambda$CDM case the suppression in the halo bispectrum is more ore less constant across the range of scale where PT gives a good fit to the measurements., revealing that difference in the shape of the underlying dark matter power spectrum are very degenerate with the bias parameters. 

\subsection{The universality of the halo bias at quadratic order}

In the context of bias modeling, universality means that the bias coefficients, as a function of mass, cosmology and redshift, can be written in terms of the peak height
$\nu \equiv \delta_{cr}/\sigma(M,z)$, defined as the ratio between the constant critical density $\delta_{cr} = 1.686$ for spherical collapse,  and $\sigma(M,z)$, the r.m.s of the linear density field smoothed at a mass scale $M$ and redshift $z$. Any dependence on the cosmological model, or redshift, is encoded in the function $\sigma(M,z)$.  This is a very strong statement, and in principle of great value for cosmological analyses, as it allows to predict, for instance, the evolution of the bias parameters with redshift. Measurements in the N-body simulations of  $\Lambda$CDM cosmologies show that bias parameters are universal functions of redshift \citep{ShethTormen1999, TinkerEtal2010, HoffmannBelGaztanaga2015, LazeyrasEtal2016, ModiCastorinaSeljak2017}.

\begin{figure}
\begin{center}
\includegraphics[width= \textwidth]{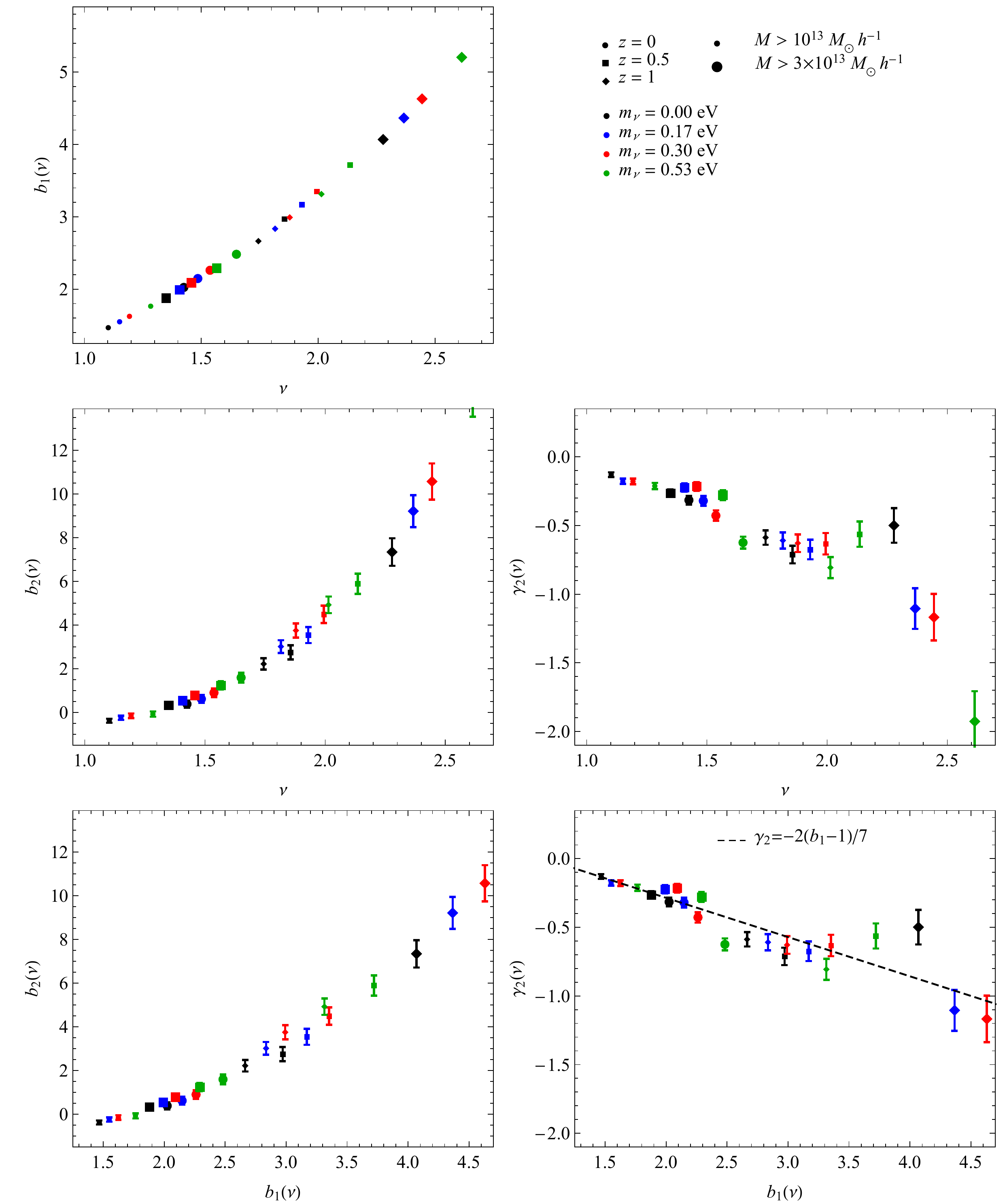}
\caption{\label{fig:bias} Upper panels: Best fit value for the bias parameters as a function the peak height for different cosmologies, redshift and halo mass threshold. Lower panels: Relation between second order bias parameter and linear bias for the same cosmological models, redshift and halo masses. }
\end{center}
\end{figure}

In \citep{CastorinaEtal2014} it was shown that the same result applies to linear bias in massive neutrinos cosmologies if the peak height is computed from the variance of the CDM field, $\nu_c = \delta_{cr}/\sigma_c$. An incorrect choice for the variance leads to strong violations of universality with both redshift and cosmology. This is just a consequence of the fact that the proper bias expansion is written in terms of the CDM field.
Given our measurements of the bispectrum and the best fit values of the bias coefficients, we are in the position to test universality beyond linear bias. Such an analysis is presented in Fig.~\ref{fig:bias}. The top panel shows the best fit value for linear bias from the power spectrum as a function of $\nu_c$. Different symbols and colors refer to different halo populations in different cosmologies, redshifts and mass thresholds. The figure agrees with \citep{CastorinaEtal2014} in showing $b_1(\nu_c)$ as a universal prediction, function of $\nu_c$ alone. The middle and bottom panel, in addition, show the universality of quadratic bias coefficients, both local and non local ones.  We find that $b_2$ and $\gamma_2$ are universal functions of cosmology and redshift if the right variable, $\nu_c$, is used. This is non-trivial check of the bias model and confirms our understanding of the clustering of halos in cosmologies with massive neutrinos.
An important consequence of universality is the existence of smooth relations between linear bias and other bias parameters. 
Such relations, if calibrated with enough accuracy, can be used to reduce to reduce the number of nuisance parameters in a cosmological analysis. In particular the Eulerian bias model assumes that non local terms are only generated by gravitational evolution, yielding   \citep{ChanScoccimarro2012, BaldaufEtal2012},
\be
\label{eq:b1g2E}
\gamma_2 = -\frac{2}{7}(b_1-1)\,,
\ee
which is assumed to be valid in all cosmological analysis of galaxy survey data \citep{SanchezEtal2017b, GilMarinEtal2017, BeutlerEtal2017, SlepianEisenstein2017}. Recently  \citep{ShethChanScoccimarro2013, CastorinaEtal2016, CastorinaParanjapeSheth2017, ModiCastorinaSeljak2017} have shown, in analytical calculations and measurements in N-body simulations of $\Lambda$CDM cosmologies, that, as a results of the fact that halo formation happens in a ellipsoidal fashion, the above equation needs to be modified to include a Lagrangian non local coefficient $\gamma_2^L$,
\be
\label{eq:b1g2L}
\gamma_2 = \gamma_2^L-\frac{2}{7}(b_1-1).
\ee

As a final application of our results on the halo bispectrum we therefore test the relations between bias parameters in cosmologies with massive neutrinos. In figure~\ref{fig:bias}, we show $b_2$, top panel, and $\gamma_2$, bottom panel, as a function of linear bias. As expected, quadratic density bias is a smooth function of $b_1$ independently of the value of the neutrino masses. This implies that existing fitting formulae for $b_2(b_1)$ as in \citep{HoffmannBelGaztanaga2017, LazeyrasEtal2016} can be used in cosmologies with massive neutrinos. Moving to non local bias, we find that the prediction of Eq. \ref{eq:b1g2E} compare reasonably well with the measurements. For high values of $\nu$ where we expect bigger deviations from Eq. \ref{eq:b1g2E} our best fit values are too noisy to say anything conclusive. We plan to return to this issue in future work, as assumptions on non local bias can affect galaxy clustering analyses that use Lagrangian \citep{ChuangEtal2017, Wang+14, VlahSeljakBaldauf2015} or Eulerian \citep{SanchezEtal2017b, GilMarinEtal2017, BeutlerEtal2017, SlepianEtal2016} perturbation theory.

\section{Conclusions}
\label{sec:conclusions}

In this work we have presented the first analysis of the matter and halo bispectrum from simulations of cosmologies with massive neutrinos described as a additional set of particles. We have measured the CDM and CDM+$\nu$ bispectrum, which we have then compared to perturbation theory predictions. Firstly we have shown, using analytical arguments, that numerical approaches including neutrinos only at linear level or through response functions could potentially predict biased bispectra on large scales.
From measurements in N-body simulations we showed that the CDM bispectrum, $B_{ccc}$ is the dominant three-point statistics, with bispectra involving one or more neutrino fields being highly suppressed. This simplifies a lot the analytical evaluation, since only $B_{ccc}$ needs to be computed beyond the tree level prediction. The perturbative calculations agrees fairly well with the N-body, most importantly at the same level it does in a standard cosmological simulation. We have shown that tree-level perturbation theory is sufficient to describe any bispectra involving one or more neutrino field, as their perturbations are highly suppressed below the free-streaming scale. 
We have also estimated non-linear neutrinos signatures in the bispectrum by looking at the reduce bispectrum, finding $<1\%$ effects for the considerably high value of the neutrino masses considered in this paper. 

We then devoted our attention to the halo bispectrum in cosmologies with massive neutrinos, the main motivation being the result in \citep{CastorinaEtal2014} that linear halo bias should be written in terms of the CDM field only. 
We extend this finding to higher order bias coefficients, showing that the halo bispectrum can be characterized by the same bias expansion is usually performed in a $\Lambda CDM$ universe. This has important consequences for universality of higher order bias parameters, which holds if written in terms of the peak height of the CDM field, $\nu_c = \delta_{cr}/\sigma_c$. This implies, for instance, that quadratic bias, $b_2$, can be written in terms of $b_1$ regardless of the value of neutrino masses.
\bigskip

\acknowledgments
The DEMNUni simulations were carried out at the Tier-0 IBM BG/Q machine, Fermi, of
the Centro Interuniversitario del Nord-Est per il Calcolo Elettronico (CINECA, Bologna,
Italy), via the five million cpu-hrs budget provided by the Italian SuperComputing Resource
Allocation (ISCRA) to the class-A proposal entitled ``The Dark Energy and Massive-Neutrino
Universe''. 
RR acknowledge support from the European Research Council
through the Darksurvey grant 614030.
CC acknowledges financial support from the European Research Council through the Darklight Advanced Research Grant
(n. 291521), and the grant MIUR PRIN 2015 ``Cosmology and Fundamental Physics: illuminating the Dark Universe with Euclid''.
EC and ES would like to thank Francisco Villaescusa-Navarro, Matteo Viel and Julien Bel for useful discussions.

\appendix

%
%
%
%

\bibliographystyle{JHEPc}
\bibliography{Bibliography}

\providecommand{\href}[2]{#2}\begingroup\raggedright\begin{thebibliography}{100}

\bibitem{PLANCK2016cp}
{Planck Collaboration}, P.~A.~R. {Ade}, N.~{Aghanim}, M.~{Arnaud},
  M.~{Ashdown}, J.~{Aumont} et~al., \emph{{Planck 2015 results. XIII.
  Cosmological parameters}},
  \href{http://dx.doi.org/10.1051/0004-6361/201525830}{\emph{\aap} {\bfseries
  594} (Sept., 2016) A13}, [\href{https://arxiv.org/abs/1502.01589}{{\ttfamily
  1502.01589}}].

\bibitem{AlamEtal2016a}
S.~{Alam}, M.~{Ata}, S.~{Bailey}, F.~{Beutler}, D.~{Bizyaev}, J.~A. {Blazek}
  et~al., \emph{{The clustering of galaxies in the completed SDSS-III Baryon
  Oscillation Spectroscopic Survey: cosmological analysis of the DR12 galaxy
  sample}}, {\emph{ArXiv e-prints} (July, 2016) },
  [\href{https://arxiv.org/abs/1607.03155}{{\ttfamily 1607.03155}}].

\bibitem{ZhaoEtal2017a}
G.-B. {Zhao}, M.~{Raveri}, L.~{Pogosian}, Y.~{Wang}, R.~G. {Crittenden}, W.~J.
  {Handley} et~al., \emph{{The clustering of galaxies in the completed SDSS-III
  Baryon Oscillation Spectroscopic Survey: Examining the observational evidence
  for dynamical dark energy}}, {\emph{ArXiv e-prints} (Jan., 2017) },
  [\href{https://arxiv.org/abs/1701.08165}{{\ttfamily 1701.08165}}].

\bibitem{LaureijsEtal2011}
R.~{Laureijs}, J.~{Amiaux}, S.~{Arduini}, J.~. {Augu{\`e}res}, J.~{Brinchmann},
  R.~{Cole} et~al., \emph{{Euclid Definition Study Report}}, {\emph{ArXiv:
  1110.3193} (Oct., 2011) }, [\href{https://arxiv.org/abs/1110.3193}{{\ttfamily
  1110.3193}}].

\bibitem{LeviEtal2013}
M.~{Levi}, C.~{Bebek}, T.~{Beers}, R.~{Blum}, R.~{Cahn}, D.~{Eisenstein}
  et~al., \emph{{The DESI Experiment, a whitepaper for Snowmass 2013}},
  {\emph{ArXiv: 1308.0847} (Aug., 2013) },
  [\href{https://arxiv.org/abs/1308.0847}{{\ttfamily 1308.0847}}].

\bibitem{LSST2017}
{LSST Science Collaboration}, P.~{Marshall}, T.~{Anguita}, F.~B. {Bianco},
  E.~C. {Bellm}, N.~{Brandt} et~al., \emph{{Science-Driven Optimization of the
  LSST Observing Strategy}}, {\emph{ArXiv e-prints} (Aug., 2017) },
  [\href{https://arxiv.org/abs/1708.04058}{{\ttfamily 1708.04058}}].

\bibitem{McDonaldEisenstein2007}
P.~{McDonald} and D.~J. {Eisenstein}, \emph{{Dark energy and curvature from a
  future baryonic acoustic oscillation survey using the Lyman-{$\alpha$}
  forest}}, \href{http://dx.doi.org/10.1103/PhysRevD.76.063009}{\emph{\prd}
  {\bfseries 76} (Sept., 2007) 063009},
  [\href{https://arxiv.org/abs/astro-ph/0607122}{{\ttfamily
  astro-ph/0607122}}].

\bibitem{EisensteinSeoWhite2007}
D.~J. {Eisenstein}, H.-J. {Seo} and M.~{White}, \emph{{On the Robustness of the
  Acoustic Scale in the Low-Redshift Clustering of Matter}},
  \href{http://dx.doi.org/10.1086/518755}{\emph{\apj} {\bfseries 664} (Aug.,
  2007) 660--674}, [\href{https://arxiv.org/abs/astro-ph/0604361}{{\ttfamily
  astro-ph/0604361}}].

\bibitem{CarboneMangilliVerde2011}
C.~{Carbone}, A.~{Mangilli} and L.~{Verde}, \emph{{Isocurvature modes and
  Baryon Acoustic Oscillations II: gains from combining CMB and Large Scale
  Structure}},
  \href{http://dx.doi.org/10.1088/1475-7516/2011/09/028}{\emph{\jcap}
  {\bfseries 9} (Sept., 2011) 028},
  [\href{https://arxiv.org/abs/1107.1211}{{\ttfamily 1107.1211}}].

\bibitem{GiusarmaEtal2011}
E.~{Giusarma}, M.~{Corsi}, M.~{Archidiacono}, R.~{de Putter}, A.~{Melchiorri},
  O.~{Mena} et~al., \emph{{Constraints on massive sterile neutrino species from
  current and future cosmological data}},
  \href{http://dx.doi.org/10.1103/PhysRevD.83.115023}{\emph{\prd} {\bfseries
  83} (June, 2011) 115023}, [\href{https://arxiv.org/abs/1102.4774}{{\ttfamily
  1102.4774}}].

\bibitem{AmendolaEtal2013}
L.~{Amendola}, S.~{Appleby}, D.~{Bacon}, T.~{Baker}, M.~{Baldi}, N.~{Bartolo}
  et~al., \emph{{Cosmology and Fundamental Physics with the Euclid Satellite}},
  \href{http://dx.doi.org/10.12942/lrr-2013-6}{\emph{Living Reviews in
  Relativity} {\bfseries 16} (Sept., 2013) },
  [\href{https://arxiv.org/abs/1206.1225}{{\ttfamily 1206.1225}}].

\bibitem{WeinbergEtal2013}
D.~H. {Weinberg}, M.~J. {Mortonson}, D.~J. {Eisenstein}, C.~{Hirata}, A.~G.
  {Riess} and E.~{Rozo}, \emph{{Observational probes of cosmic acceleration}},
  \href{http://dx.doi.org/10.1016/j.physrep.2013.05.001}{\emph{\physrep}
  {\bfseries 530} (Sept., 2013) 87--255},
  [\href{https://arxiv.org/abs/1201.2434}{{\ttfamily 1201.2434}}].

\bibitem{AudrenEtal2013}
B.~Audren, J.~Lesgourgues, S.~{Bird}, M.~G. {Haehnelt} and M.~Viel,
  \emph{{Neutrino masses and cosmological parameters from a Euclid-like survey:
  Markov Chain Monte Carlo forecasts including theoretical errors}},
  \href{http://dx.doi.org/10.1088/1475-7516/2013/01/026}{\emph{\jcap}
  {\bfseries 1} (Jan., 2013) 26},
  [\href{https://arxiv.org/abs/1210.2194}{{\ttfamily 1210.2194}}].

\bibitem{JoyceEtal2015}
A.~{Joyce}, B.~{Jain}, J.~{Khoury} and M.~{Trodden}, \emph{{Beyond the
  cosmological standard model}},
  \href{http://dx.doi.org/10.1016/j.physrep.2014.12.002}{\emph{\physrep}
  {\bfseries 568} (Mar., 2015) 1--98},
  [\href{https://arxiv.org/abs/1407.0059}{{\ttfamily 1407.0059}}].

\bibitem{BaldaufEtal2016}
T.~{Baldauf}, M.~{Mirbabayi}, M.~{Simonovi{\'c}} and M.~{Zaldarriaga},
  \emph{{LSS constraints with controlled theoretical uncertainties}},
  {\emph{ArXiv e-prints} (Feb., 2016) },
  [\href{https://arxiv.org/abs/1602.00674}{{\ttfamily 1602.00674}}].

\bibitem{DesjacquesJeongSchmidt2016}
V.~{Desjacques}, D.~{Jeong} and F.~{Schmidt}, \emph{{Large-Scale Galaxy Bias}},
  {\emph{ArXiv e-prints} (Nov., 2016) },
  [\href{https://arxiv.org/abs/1611.09787}{{\ttfamily 1611.09787}}].

\bibitem{Koyama2016}
K.~{Koyama}, \emph{{Cosmological tests of modified gravity}},
  \href{http://dx.doi.org/10.1088/0034-4885/79/4/046902}{\emph{Reports on
  Progress in Physics} {\bfseries 79} (Apr., 2016) 046902},
  [\href{https://arxiv.org/abs/1504.04623}{{\ttfamily 1504.04623}}].

\bibitem{AlonsoEtal2017}
D.~{Alonso}, E.~{Bellini}, P.~G. {Ferreira} and M.~{Zumalac{\'a}rregui},
  \emph{{Observational future of cosmological scalar-tensor theories}},
  \href{http://dx.doi.org/10.1103/PhysRevD.95.063502}{\emph{\prd} {\bfseries
  95} (Mar., 2017) 063502}, [\href{https://arxiv.org/abs/1610.09290}{{\ttfamily
  1610.09290}}].

\bibitem{LesgourguesPastor2006}
J.~{Lesgourgues} and S.~{Pastor}, \emph{{Massive neutrinos and cosmology}},
  \href{http://dx.doi.org/10.1016/j.physrep.2006.04.001}{\emph{\physrep}
  {\bfseries 429} (July, 2006) 307--379},
  [\href{https://arxiv.org/abs/astro-ph/0603494}{{\ttfamily
  astro-ph/0603494}}].

\bibitem{Lesgourgues2013}
J.~{Lesgourgues}, G.~{Mangano}, G.~{Miele} and S.~{Pastor}, \emph{{Neutrino
  Cosmology}}.
\newblock Cambridge, UK: Cambridge University Press, Feb., 2013.

\bibitem{PalanqueDelabrouilleEtal2015}
N.~{Palanque-Delabrouille}, C.~{Y{\`e}che}, J.~{Baur}, C.~{Magneville},
  G.~{Rossi}, J.~{Lesgourgues} et~al., \emph{{Neutrino masses and cosmology
  with Lyman-alpha forest power spectrum}},
  \href{http://dx.doi.org/10.1088/1475-7516/2015/11/011}{\emph{\jcap}
  {\bfseries 11} (Nov., 2015) 011},
  [\href{https://arxiv.org/abs/1506.05976}{{\ttfamily 1506.05976}}].

\bibitem{PLANCK2016SZClusters}
{Planck Collaboration}, P.~A.~R. {Ade}, N.~{Aghanim}, M.~{Arnaud},
  M.~{Ashdown}, J.~{Aumont} et~al., \emph{{Planck 2015 results. XXIV. Cosmology
  from Sunyaev-Zeldovich cluster counts}},
  \href{http://dx.doi.org/10.1051/0004-6361/201525833}{\emph{\aap} {\bfseries
  594} (Sept., 2016) A24}, [\href{https://arxiv.org/abs/1502.01597}{{\ttfamily
  1502.01597}}].

\bibitem{CuestaNiroVerde2016}
A.~J. {Cuesta}, V.~{Niro} and L.~{Verde}, \emph{{Neutrino mass limits: Robust
  information from the power spectrum of galaxy surveys}},
  \href{http://dx.doi.org/10.1016/j.dark.2016.04.005}{\emph{Physics of the Dark
  Universe} {\bfseries 13} (Sept., 2016) 77--86},
  [\href{https://arxiv.org/abs/1511.05983}{{\ttfamily 1511.05983}}].

\bibitem{VagnozziEtal2017}
S.~{Vagnozzi}, E.~{Giusarma}, O.~{Mena}, K.~{Freese}, M.~{Gerbino}, S.~{Ho}
  et~al., \emph{{Unveiling $\nu$ secrets with cosmological data: neutrino
  masses and mass hierarchy}}, {\emph{ArXiv e-prints} (Jan., 2017) },
  [\href{https://arxiv.org/abs/1701.08172}{{\ttfamily 1701.08172}}].

\bibitem{ArchidiaconoEtal2017}
M.~{Archidiacono}, T.~{Brinckmann}, J.~{Lesgourgues} and V.~{Poulin},
  \emph{{Physical effects involved in the measurements of neutrino masses with
  future cosmological data}},
  \href{http://dx.doi.org/10.1088/1475-7516/2017/02/052}{\emph{\jcap}
  {\bfseries 2} (Feb., 2017) 052},
  [\href{https://arxiv.org/abs/1610.09852}{{\ttfamily 1610.09852}}].

\bibitem{SaitoTakadaTaruya2008}
S.~Saito, M.~Takada and A.~Taruya, \emph{{Impact of Massive Neutrinos on the
  Nonlinear Matter Power Spectrum}},
  \href{http://dx.doi.org/10.1103/PhysRevLett.100.191301}{\emph{Physical Review
  Letters} {\bfseries 100} (May, 2008) 191301},
  [\href{https://arxiv.org/abs/0801.0607}{{\ttfamily 0801.0607}}].

\bibitem{Wong2008}
Y.~Y.~Y. Wong, \emph{{Higher order corrections to the large scale matter power
  spectrum in the presence of massive neutrinos}},
  \href{http://dx.doi.org/10.1088/1475-7516/2008/10/035}{\emph{\jcap}
  {\bfseries 10} (Oct., 2008) 35},
  [\href{https://arxiv.org/abs/0809.0693}{{\ttfamily 0809.0693}}].

\bibitem{LesgourguesEtal2009}
J.~Lesgourgues, S.~Matarrese, M.~Pietroni and A.~Riotto, \emph{Non-linear power
  spectrum including massive neutrinos: the time-rg flow approach},
  \href{http://dx.doi.org/10.1088/1475-7516/2009/06/017}{\emph{\jcap}
  {\bfseries 6} (June, 2009) 17}, [\href{https://arxiv.org/abs/arXiv: 0901.4550
  [astro-ph.CO]}{{\ttfamily arXiv: 0901.4550 [astro-ph.CO]}}].

\bibitem{UpadhyeEtal2014}
A.~{Upadhye}, R.~{Biswas}, A.~{Pope}, K.~{Heitmann}, S.~{Habib}, H.~{Finkel}
  et~al., \emph{{Large-scale structure formation with massive neutrinos and
  dynamical dark energy}},
  \href{http://dx.doi.org/10.1103/PhysRevD.89.103515}{\emph{\prd} {\bfseries
  89} (May, 2014) 103515}, [\href{https://arxiv.org/abs/1309.5872}{{\ttfamily
  1309.5872}}].

\bibitem{BlasEtal2014}
D.~{Blas}, M.~{Garny}, T.~{Konstandin} and J.~{Lesgourgues}, \emph{{Structure
  formation with massive neutrinos: going beyond linear theory}},
  \href{http://dx.doi.org/10.1088/1475-7516/2014/11/039}{\emph{\jcap}
  {\bfseries 11} (Nov., 2014) 039},
  [\href{https://arxiv.org/abs/1408.2995}{{\ttfamily 1408.2995}}].

\bibitem{DupuyBernardeau2014}
H.~Dupuy and F.~Bernardeau, \emph{{Describing massive neutrinos in cosmology as
  a collection of independent flows}},
  \href{http://dx.doi.org/10.1088/1475-7516/2014/01/030}{\emph{\jcap}
  {\bfseries 1} (Jan., 2014) 30},
  [\href{https://arxiv.org/abs/1311.5487}{{\ttfamily 1311.5487}}].

\bibitem{DupuyBernardeau2015}
H.~{Dupuy} and F.~Bernardeau, \emph{{On the importance of nonlinear couplings
  in large-scale neutrino streams}},
  \href{http://dx.doi.org/10.1088/1475-7516/2015/08/053}{\emph{\jcap}
  {\bfseries 8} (Aug., 2015) 053},
  [\href{https://arxiv.org/abs/1503.05707}{{\ttfamily 1503.05707}}].

\bibitem{CastorinaEtal2015}
E.~{Castorina}, C.~{Carbone}, J.~{Bel}, E.~{Sefusatti} and K.~{Dolag},
  \emph{{DEMNUni: the clustering of large-scale structures in the presence of
  massive neutrinos}},
  \href{http://dx.doi.org/10.1088/1475-7516/2015/07/043}{\emph{\jcap}
  {\bfseries 7} (July, 2015) 043},
  [\href{https://arxiv.org/abs/1505.07148}{{\ttfamily 1505.07148}}].

\bibitem{FuhrerWong2015}
F.~{F{\"u}hrer} and Y.~Y.~Y. {Wong}, \emph{{Higher-order massive neutrino
  perturbations in large-scale structure}},
  \href{http://dx.doi.org/10.1088/1475-7516/2015/03/046}{\emph{\jcap}
  {\bfseries 3} (Mar., 2015) 046},
  [\href{https://arxiv.org/abs/1412.2764}{{\ttfamily 1412.2764}}].

\bibitem{UpadhyeEtal2016}
A.~{Upadhye}, J.~{Kwan}, A.~{Pope}, K.~{Heitmann}, S.~{Habib}, H.~{Finkel}
  et~al., \emph{{Redshift-space distortions in massive neutrino and evolving
  dark energy cosmologies}},
  \href{http://dx.doi.org/10.1103/PhysRevD.93.063515}{\emph{\prd} {\bfseries
  93} (Mar., 2016) 063515}, [\href{https://arxiv.org/abs/1506.07526}{{\ttfamily
  1506.07526}}].

\bibitem{ArchidiaconoHannestad2016}
M.~{Archidiacono} and S.~{Hannestad}, \emph{{Efficient calculation of
  cosmological neutrino clustering in the non-linear regime}},
  \href{http://dx.doi.org/10.1088/1475-7516/2016/06/018}{\emph{\jcap}
  {\bfseries 6} (June, 2016) 018},
  [\href{https://arxiv.org/abs/1510.02907}{{\ttfamily 1510.02907}}].

\bibitem{LeviVlah2016}
M.~{Levi} and Z.~Vlah, \emph{{Massive neutrinos in nonlinear large scale
  structure: A consistent perturbation theory}}, {\emph{ArXiv e-prints} (May,
  2016) }, [\href{https://arxiv.org/abs/1605.09417}{{\ttfamily 1605.09417}}].

\bibitem{LiuEtal2017A}
J.~{Liu}, S.~{Bird}, J.~M. {Zorrilla Matilla}, J.~C. {Hill}, Z.~{Haiman}, M.~S.
  {Madhavacheril} et~al., \emph{{MassiveNuS: Cosmological Massive Neutrino
  Simulations}}, {\emph{ArXiv e-prints} (Nov., 2017) },
  [\href{https://arxiv.org/abs/1711.10524}{{\ttfamily 1711.10524}}].

\bibitem{VillaescusaEtal2014}
F.~{Villaescusa-Navarro}, F.~{Marulli}, M.~{Viel}, E.~{Branchini},
  E.~{Castorina}, E.~{Sefusatti} et~al., \emph{{Cosmology with massive
  neutrinos I: towards a realistic modeling of the relation between matter,
  haloes and galaxies}},
  \href{http://dx.doi.org/10.1088/1475-7516/2014/03/011}{\emph{\jcap}
  {\bfseries 3} (Mar., 2014) 11},
  [\href{https://arxiv.org/abs/1311.0866}{{\ttfamily 1311.0866}}].

\bibitem{LiuEtal2017}
Y.~{Liu}, Y.~{Liang}, H.-R. {Yu}, C.~{Zhao}, J.~{Qin} and T.-J. {Zhang},
  \emph{{Baryon Acoustic Oscillation detections from the clustering of massive
  halos and different density region tracers in TianNu simulation}},
  {\emph{ArXiv e-prints} (Dec., 2017) },
  [\href{https://arxiv.org/abs/1712.01002}{{\ttfamily 1712.01002}}].

\bibitem{IchikiTakada2012}
K.~Ichiki and M.~Takada, \emph{{Impact of massive neutrinos on the abundance of
  massive clusters}},
  \href{http://dx.doi.org/10.1103/PhysRevD.85.063521}{\emph{\prd} {\bfseries
  85} (Mar., 2012) 063521}, [\href{https://arxiv.org/abs/1108.4688}{{\ttfamily
  1108.4688}}].

\bibitem{CastorinaEtal2014}
E.~Castorina, E.~Sefusatti, R.~K. Sheth, F.~Villaescusa-Navarro and M.~Viel,
  \emph{{Cosmology with massive neutrinos II: on the universality of the halo
  mass function and bias}},
  \href{http://dx.doi.org/10.1088/1475-7516/2014/02/049}{\emph{\jcap}
  {\bfseries 2} (Feb., 2014) 49},
  [\href{https://arxiv.org/abs/1311.1212}{{\ttfamily 1311.1212}}].

\bibitem{CostanziEtal2013B}
M.~Costanzi, F.~Villaescusa-Navarro, M.~Viel, J.-Q. Xia, S.~Borgani,
  E.~Castorina et~al., \emph{{Cosmology with massive neutrinos III: the halo
  mass function and an application to galaxy clusters}},
  \href{http://dx.doi.org/10.1088/1475-7516/2013/12/012}{\emph{\jcap}
  {\bfseries 12} (Dec., 2013) 12},
  [\href{https://arxiv.org/abs/1311.1514}{{\ttfamily 1311.1514}}].

\bibitem{LoVerde2014B}
M.~{LoVerde}, \emph{{Spherical collapse in {$\nu$}{$\Lambda$}CDM}},
  \href{http://dx.doi.org/10.1103/PhysRevD.90.083518}{\emph{\prd} {\bfseries
  90} (Oct., 2014) 083518}, [\href{https://arxiv.org/abs/1405.4858}{{\ttfamily
  1405.4858}}].

\bibitem{BiagettiEtal2014}
M.~{Biagetti}, K.~C. {Chan}, V.~{Desjacques} and A.~{Paranjape},
  \emph{{Measuring non-local Lagrangian peak bias}},
  \href{http://dx.doi.org/10.1093/mnras/stu680}{\emph{\mnras} {\bfseries 441}
  (June, 2014) 1457--1467}, [\href{https://arxiv.org/abs/1310.1401}{{\ttfamily
  1310.1401}}].

\bibitem{LoVerdeZaldarriaga2014}
M.~LoVerde and M.~Zaldarriaga, \emph{{Neutrino clustering around spherical dark
  matter halos}},
  \href{http://dx.doi.org/10.1103/PhysRevD.89.063502}{\emph{\prd} {\bfseries
  89} (Mar., 2014) 063502}, [\href{https://arxiv.org/abs/1310.6459}{{\ttfamily
  1310.6459}}].

\bibitem{LoVerde2016}
M.~{LoVerde}, \emph{{Neutrino mass without cosmic variance}},
  \href{http://dx.doi.org/10.1103/PhysRevD.93.103526}{\emph{\prd} {\bfseries
  93} (May, 2016) 103526}, [\href{https://arxiv.org/abs/1602.08108}{{\ttfamily
  1602.08108}}].

\bibitem{ChiangEtal2017}
C.-T. {Chiang}, W.~{Hu}, Y.~{Li} and M.~{LoVerde}, \emph{{Scale-dependent bias
  and bispectrum in neutrino separate universe simulations}}, {\emph{ArXiv
  e-prints} (Oct., 2017) }, [\href{https://arxiv.org/abs/1710.01310}{{\ttfamily
  1710.01310}}].

\bibitem{MassaraEtal2015}
E.~{Massara}, F.~{Villaescusa-Navarro}, M.~{Viel} and P.~M. {Sutter},
  \emph{{Voids in massive neutrino cosmologies}},
  \href{http://dx.doi.org/10.1088/1475-7516/2015/11/018}{\emph{\jcap}
  {\bfseries 11} (Nov., 2015) 018},
  [\href{https://arxiv.org/abs/1506.03088}{{\ttfamily 1506.03088}}].

\bibitem{ZhuEtal2014}
H.-M. {Zhu}, U.-L. {Pen}, X.~{Chen}, D.~{Inman} and Y.~{Yu}, \emph{{Measurement
  of Neutrino Masses from Relative Velocities}},
  \href{http://dx.doi.org/10.1103/PhysRevLett.113.131301}{\emph{Physical Review
  Letters} {\bfseries 113} (Sept., 2014) 131301},
  [\href{https://arxiv.org/abs/1311.3422}{{\ttfamily 1311.3422}}].

\bibitem{OkoliEtal2016}
C.~{Okoli}, M.~I. {Scrimgeour}, N.~{Afshordi} and M.~J. {Hudson},
  \emph{{Dynamical friction in the primordial neutrino sea}}, {\emph{ArXiv
  e-prints} (Nov., 2016) }, [\href{https://arxiv.org/abs/1611.04589}{{\ttfamily
  1611.04589}}].

\bibitem{SenatoreZaldarriaga2017}
L.~{Senatore} and M.~{Zaldarriaga}, \emph{{The Effective Field Theory of
  Large-Scale Structure in the presence of Massive Neutrinos}}, {\emph{ArXiv
  e-prints} (July, 2017) }, [\href{https://arxiv.org/abs/1707.04698}{{\ttfamily
  1707.04698}}].

\bibitem{ShojiKomatsu2010}
M.~{Shoji} and E.~Komatsu, \emph{{Massive neutrinos in cosmology: Analytic
  solutions and fluid approximation}},
  \href{http://dx.doi.org/10.1103/PhysRevD.81.123516}{\emph{\prd} {\bfseries
  81} (June, 2010) 123516}.

\bibitem{ChanScoccimarroSheth2012}
K.~C. Chan, R.~Scoccimarro and R.~K. Sheth, \emph{{Gravity and large-scale
  nonlocal bias}},
  \href{http://dx.doi.org/10.1103/PhysRevD.85.083509}{\emph{\prd} {\bfseries
  85} (Apr., 2012) 083509}, [\href{https://arxiv.org/abs/1201.3614}{{\ttfamily
  1201.3614}}].

\bibitem{BaldaufEtal2012}
T.~{Baldauf}, U.~{Seljak}, V.~{Desjacques} and P.~{McDonald}, \emph{{Evidence
  for quadratic tidal tensor bias from the halo bispectrum}},
  \href{http://dx.doi.org/10.1103/PhysRevD.86.083540}{\emph{\prd} {\bfseries
  86} (Oct., 2012) 083540}, [\href{https://arxiv.org/abs/1201.4827}{{\ttfamily
  1201.4827}}].

\bibitem{SefusattiEtal2006}
E.~Sefusatti, M.~Crocce, S.~Pueblas and R.~Scoccimarro, \emph{Cosmology and the
  bispectrum}, \href{http://dx.doi.org/10.1103/PhysRevD.74.023522}{\emph{\prd}
  {\bfseries 74} (July, 2006) 023522}, [\href{https://arxiv.org/abs/arXiv:
  astro-ph/0604505}{{\ttfamily arXiv: astro-ph/0604505}}].

\bibitem{SlepianEtal2016}
Z.~{Slepian}, D.~J. {Eisenstein}, J.~R. {Brownstein}, C.-H. {Chuang},
  H.~{Gil-Mar{\'{\i}}n}, S.~{Ho} et~al., \emph{{Detection of Baryon Acoustic
  Oscillation Features in the Large-Scale 3-Point Correlation Function of SDSS
  BOSS DR12 CMASS Galaxies}}, {\emph{ArXiv e-prints} (July, 2016) },
  [\href{https://arxiv.org/abs/1607.06097}{{\ttfamily 1607.06097}}].

\bibitem{GilMarinEtal2017}
H.~{Gil-Mar{\'{\i}}n}, W.~J. {Percival}, L.~{Verde}, J.~R. {Brownstein}, C.-H.
  {Chuang}, F.-S. {Kitaura} et~al., \emph{{The clustering of galaxies in the
  SDSS-III Baryon Oscillation Spectroscopic Survey: RSD measurement from the
  power spectrum and bispectrum of the DR12 BOSS galaxies}},
  \href{http://dx.doi.org/10.1093/mnras/stw2679}{\emph{\mnras} {\bfseries 465}
  (Feb., 2017) 1757--1788}, [\href{https://arxiv.org/abs/1606.00439}{{\ttfamily
  1606.00439}}].

\bibitem{HuEisensteinTegmark1998}
W.~Hu, D.~J. Eisenstein and M.~Tegmark, \emph{{Weighing Neutrinos with Galaxy
  Surveys}},
  \href{http://dx.doi.org/10.1103/PhysRevLett.80.5255}{\emph{Physical Review
  Letters} {\bfseries 80} (June, 1998) 5255--5258},
  [\href{https://arxiv.org/abs/astro-ph/9712057}{{\ttfamily
  astro-ph/9712057}}].

\bibitem{BrandbygeEtal2008}
J.~{Brandbyge}, S.~{Hannestad}, T.~{Haugb{\o}lle} and B.~{Thomsen}, \emph{{The
  effect of thermal neutrino motion on the non-linear cosmological matter power
  spectrum}},
  \href{http://dx.doi.org/10.1088/1475-7516/2008/08/020}{\emph{\jcap}
  {\bfseries 8} (Aug., 2008) 020},
  [\href{https://arxiv.org/abs/0802.3700}{{\ttfamily 0802.3700}}].

\bibitem{VielHaehneltSpringel2010}
M.~{Viel}, M.~G. {Haehnelt} and V.~{Springel}, \emph{{The effect of neutrinos
  on the matter distribution as probed by the intergalactic medium}},
  \href{http://dx.doi.org/10.1088/1475-7516/2010/06/015}{\emph{\jcap}
  {\bfseries 6} (June, 2010) 15},
  [\href{https://arxiv.org/abs/1003.2422}{{\ttfamily 1003.2422}}].

\bibitem{BirdVielHaehnelt2012}
S.~Bird, M.~Viel and M.~G. Haehnelt, \emph{{Massive neutrinos and the
  non-linear matter power spectrum}},
  \href{http://dx.doi.org/10.1111/j.1365-2966.2011.20222.x}{\emph{\mnras}
  {\bfseries 420} (Mar., 2012) 2551--2561},
  [\href{https://arxiv.org/abs/1109.4416}{{\ttfamily 1109.4416}}].

\bibitem{WagnerVerdeJimenez2012}
C.~{Wagner}, L.~{Verde} and R.~{Jimenez}, \emph{{Effects of the Neutrino Mass
  Splitting on the Nonlinear Matter Power Spectrum}},
  \href{http://dx.doi.org/10.1088/2041-8205/752/2/L31}{\emph{\apjl} {\bfseries
  752} (June, 2012) L31}, [\href{https://arxiv.org/abs/1203.5342}{{\ttfamily
  1203.5342}}].

\bibitem{InmanEtal2015}
D.~{Inman}, J.~D. {Emberson}, U.-L. {Pen}, A.~{Farchi}, H.-R. {Yu} and
  J.~{Harnois-D{\'e}raps}, \emph{{Precision reconstruction of the cold dark
  matter-neutrino relative velocity from N -body simulations}},
  \href{http://dx.doi.org/10.1103/PhysRevD.92.023502}{\emph{\prd} {\bfseries
  92} (July, 2015) 023502}, [\href{https://arxiv.org/abs/1503.07480}{{\ttfamily
  1503.07480}}].

\bibitem{YuEtal2017}
H.-R. {Yu}, J.~D. {Emberson}, D.~{Inman}, T.-J. {Zhang}, U.-L. {Pen},
  J.~{Harnois-D{\'e}raps} et~al., \emph{{Differential neutrino condensation
  onto cosmic structure}},
  \href{http://dx.doi.org/10.1038/s41550-017-0143}{\emph{Nature Astronomy}
  {\bfseries 1} (July, 2017) 0143},
  [\href{https://arxiv.org/abs/1609.08968}{{\ttfamily 1609.08968}}].

\bibitem{BernardeauEtal2002}
F.~Bernardeau, S.~Colombi, E.~Gazta{\~n}aga and R.~Scoccimarro,
  \emph{Large-scale structure of the universe and cosmological perturbation
  theory}, {\emph{\physrep} {\bfseries 367} (Sept., 2002) 1--3},
  [\href{https://arxiv.org/abs/astro-ph/0112551}{{\ttfamily
  astro-ph/0112551}}].

\bibitem{CarbonePetkovaDolag2016}
C.~{Carbone}, M.~{Petkova} and K.~{Dolag}, \emph{{DEMNUni: ISW, Rees-Sciama,
  and weak-lensing in the presence of massive neutrinos}},
  \href{http://dx.doi.org/10.1088/1475-7516/2016/07/034}{\emph{\jcap}
  {\bfseries 7} (July, 2016) 034},
  [\href{https://arxiv.org/abs/1605.02024}{{\ttfamily 1605.02024}}].

\bibitem{PLANCK2013parameters}
{Planck Collaboration}, P.~A.~R. {Ade}, N.~{Aghanim}, C.~{Armitage-Caplan},
  M.~{Arnaud}, M.~{Ashdown} et~al., \emph{{Planck 2013 results. XVI.
  Cosmological parameters}},
  \href{http://dx.doi.org/10.1051/0004-6361/201321591}{\emph{\aap} {\bfseries
  571} (Nov., 2014) A16}, [\href{https://arxiv.org/abs/1303.5076}{{\ttfamily
  1303.5076}}].

\bibitem{SpringelEtal2005}
V.~{Springel}, S.~D.~M. {White}, A.~{Jenkins}, C.~S. {Frenk}, N.~{Yoshida},
  L.~{Gao} et~al., \emph{{Simulations of the formation, evolution and
  clustering of galaxies and quasars}},
  \href{http://dx.doi.org/10.1038/nature03597}{\emph{\nat} {\bfseries 435}
  (June, 2005) 629--636},
  [\href{https://arxiv.org/abs/arXiv:astro-ph/0504097}{{\ttfamily
  arXiv:astro-ph/0504097}}].

\bibitem{SpringelYoshidaWhite2001}
V.~{Springel}, N.~{Yoshida} and S.~D.~M. {White}, \emph{{GADGET: a code for
  collisionless and gasdynamical cosmological simulations}},
  \href{http://dx.doi.org/10.1016/S1384-1076(01)00042-2}{\emph{\na} {\bfseries
  6} (Apr., 2001) 79--117},
  [\href{https://arxiv.org/abs/astro-ph/0003162}{{\ttfamily
  astro-ph/0003162}}].

\bibitem{DolagEtal2009}
K.~{Dolag}, S.~{Borgani}, G.~{Murante} and V.~{Springel}, \emph{{Substructures
  in hydrodynamical cluster simulations}},
  \href{http://dx.doi.org/10.1111/j.1365-2966.2009.15034.x}{\emph{\mnras}
  {\bfseries 399} (Oct., 2009) 497--514},
  [\href{https://arxiv.org/abs/0808.3401}{{\ttfamily 0808.3401}}].

\bibitem{DavisEtal1985}
M.~Davis, G.~P. Efstathiou, C.~S. Frenk and S.~D.~M. White, \emph{{The
  evolution of large-scale structure in a universe dominated by cold dark
  matter}}, \href{http://dx.doi.org/10.1086/163168}{\emph{\apj} {\bfseries 292}
  (May, 1985) 371--394}.

\bibitem{Scoccimarro2015}
R.~{Scoccimarro}, \emph{{Fast estimators for redshift-space clustering}},
  \href{http://dx.doi.org/10.1103/PhysRevD.92.083532}{\emph{\prd} {\bfseries
  92} (Oct., 2015) 083532}, [\href{https://arxiv.org/abs/1506.02729}{{\ttfamily
  1506.02729}}].

\bibitem{SefusattiEtal2016}
E.~{Sefusatti}, M.~{Crocce}, R.~{Scoccimarro} and H.~M.~P. {Couchman},
  \emph{{Accurate estimators of correlation functions in Fourier space}},
  \href{http://dx.doi.org/10.1093/mnras/stw1229}{\emph{\mnras} {\bfseries 460}
  (Aug., 2016) 3624--3636}, [\href{https://arxiv.org/abs/1512.07295}{{\ttfamily
  1512.07295}}].

\bibitem{Scoccimarro2000B}
R.~Scoccimarro, \emph{The bispectrum: From theory to observations},
  \href{http://dx.doi.org/10.1086/317248}{\emph{\apj} {\bfseries 544} (Dec.,
  2000) 597--615}, [\href{https://arxiv.org/abs/astro-ph/0004086}{{\ttfamily
  astro-ph/0004086}}].

\bibitem{Sefusatti2009}
E.~Sefusatti, \emph{One-loop perturbative corrections to the matter and galaxy
  bispectrum with non-gaussian initial conditions},
  \href{http://dx.doi.org/10.1103/PhysRevD.80.123002}{\emph{\prd} {\bfseries
  80} (Dec., 2009) 123002}, [\href{https://arxiv.org/abs/0905.0717}{{\ttfamily
  0905.0717}}].

\bibitem{SaitoTakadaTaruya2009}
S.~Saito, M.~Takada and A.~Taruya, \emph{{Nonlinear power spectrum in the
  presence of massive neutrinos: Perturbation theory approach, galaxy bias, and
  parameter forecasts}},
  \href{http://dx.doi.org/10.1103/PhysRevD.80.083528}{\emph{\prd} {\bfseries
  80} (Oct., 2009) 083528}, [\href{https://arxiv.org/abs/0907.2922}{{\ttfamily
  0907.2922}}].

\bibitem{ScoccimarroEtal1998}
R.~Scoccimarro, S.~Colombi, J.~N. Fry, J.~A. Frieman, E.~Hivon and A.~Melott,
  \emph{Nonlinear evolution of the bispectrum of cosmological perturbations},
  \href{http://dx.doi.org/10.1086/305399}{\emph{\apj} {\bfseries 496} (Mar.,
  1998) 586}, [\href{https://arxiv.org/abs/astro-ph/9704075}{{\ttfamily
  astro-ph/9704075}}].

\bibitem{SefusattiCrocceDesjacques2010}
E.~{Sefusatti}, M.~{Crocce} and V.~{Desjacques}, \emph{{The matter bispectrum
  in N-body simulations with non-Gaussian initial conditions}},
  \href{http://dx.doi.org/10.1111/j.1365-2966.2010.16723.x}{\emph{\mnras}
  {\bfseries 406} (Aug., 2010) 1014--1028},
  [\href{https://arxiv.org/abs/1003.0007}{{\ttfamily 1003.0007}}].

\bibitem{CarrascoHertzbergSenatore2012}
J.~J.~M. {Carrasco}, M.~P. {Hertzberg} and L.~{Senatore}, \emph{{The effective
  field theory of cosmological large scale structures}},
  \href{http://dx.doi.org/10.1007/JHEP09(2012)082}{\emph{Journal of High Energy
  Physics} {\bfseries 9} (Sept., 2012) 82},
  [\href{https://arxiv.org/abs/1206.2926}{{\ttfamily 1206.2926}}].

\bibitem{AnguloEtal2015B}
R.~E. {Angulo}, S.~{Foreman}, M.~{Schmittfull} and L.~{Senatore}, \emph{{The
  one-loop matter bispectrum in the Effective Field Theory of Large Scale
  Structures}},
  \href{http://dx.doi.org/10.1088/1475-7516/2015/10/039}{\emph{\jcap}
  {\bfseries 10} (Oct., 2015) 039},
  [\href{https://arxiv.org/abs/1406.4143}{{\ttfamily 1406.4143}}].

\bibitem{FryGaztanaga1993}
J.~N. Fry and E.~Gazta{\~n}aga, \emph{Biasing and hierarchical statistics in
  large-scale structure}, \href{http://dx.doi.org/10.1086/173015}{\emph{\apj}
  {\bfseries 413} (Aug., 1993) 447--452},
  [\href{https://arxiv.org/abs/astro-ph/9302009}{{\ttfamily
  astro-ph/9302009}}].

\bibitem{PollackSmithPorciani2012}
J.~E. Pollack, R.~E. Smith and C.~Porciani, \emph{{Modelling large-scale halo
  bias using the bispectrum}},
  \href{http://dx.doi.org/10.1111/j.1365-2966.2011.20279.x}{\emph{\mnras}
  {\bfseries 420} (Mar., 2012) 3469--3489},
  [\href{https://arxiv.org/abs/1109.3458}{{\ttfamily 1109.3458}}].

\bibitem{ShethChanScoccimarro2013}
R.~K. {Sheth}, K.~C. {Chan} and R.~{Scoccimarro}, \emph{{Nonlocal Lagrangian
  bias}}, \href{http://dx.doi.org/10.1103/PhysRevD.87.083002}{\emph{\prd}
  {\bfseries 87} (Apr., 2013) 083002},
  [\href{https://arxiv.org/abs/1207.7117}{{\ttfamily 1207.7117}}].

\bibitem{EFT}
J.~J.~M. {Carrasco}, M.~P. {Hertzberg} and L.~{Senatore}, \emph{{The effective
  field theory of cosmological large scale structures}},
  \href{http://dx.doi.org/10.1007/JHEP09(2012)082}{\emph{Journal of High Energy
  Physics} {\bfseries 9} (Sept., 2012) 82},
  [\href{https://arxiv.org/abs/1206.2926}{{\ttfamily 1206.2926}}].

\bibitem{Desjacques}
V.~{Desjacques}, M.~{Crocce}, R.~{Scoccimarro} and R.~K. {Sheth},
  \emph{{Modeling scale-dependent bias on the baryonic acoustic scale with the
  statistics of peaks of Gaussian random fields}},
  \href{http://dx.doi.org/10.1103/PhysRevD.82.103529}{\emph{\prd} {\bfseries
  82} (Nov., 2010) 103529}, [\href{https://arxiv.org/abs/1009.3449}{{\ttfamily
  1009.3449}}].

\bibitem{PressEtal1992}
W.~H. {Press}, S.~A. {Teukolsky}, W.~T. {Vetterling} and B.~P. {Flannery},
  \emph{{Numerical recipes in FORTRAN. The art of scientific computing}}.
\newblock Cambridge: University Press, 1992.

\bibitem{TegmarkTaylorHeavens1997}
M.~Tegmark, A.~N. Taylor and A.~F. Heavens, \emph{Karhunen-loeve eigenvalue
  problems in cosmology: How should we tackle large data sets?},
  \href{http://dx.doi.org/10.1086/303939}{\emph{\apj} {\bfseries 480} (May,
  1997) 22}, [\href{https://arxiv.org/abs/arXiv:astro-ph/9603021}{{\ttfamily
  arXiv:astro-ph/9603021}}].

\bibitem{ChanScoccimarro2012}
K.~C. {Chan} and R.~{Scoccimarro}, \emph{{Halo sampling, local bias, and loop
  corrections}},
  \href{http://dx.doi.org/10.1103/PhysRevD.86.103519}{\emph{\prd} {\bfseries
  86} (Nov., 2012) 103519}, [\href{https://arxiv.org/abs/1204.5770}{{\ttfamily
  1204.5770}}].

\bibitem{ShethTormen1999}
R.~K. Sheth and G.~Tormen, \emph{Large-scale bias and the peak background
  split}, {\emph{\mnras} {\bfseries 308} (Sept., 1999) 119--126},
  [\href{https://arxiv.org/abs/astro-ph/9901122}{{\ttfamily
  astro-ph/9901122}}].

\bibitem{TinkerEtal2010}
J.~L. {Tinker}, B.~E. {Robertson}, A.~V. {Kravtsov}, A.~{Klypin}, M.~S.
  {Warren}, G.~{Yepes} et~al., \emph{{The Large-scale Bias of Dark Matter
  Halos: Numerical Calibration and Model Tests}},
  \href{http://dx.doi.org/10.1088/0004-637X/724/2/878}{\emph{\apj} {\bfseries
  724} (Dec., 2010) 878--886},
  [\href{https://arxiv.org/abs/1001.3162}{{\ttfamily 1001.3162}}].

\bibitem{HoffmannBelGaztanaga2015}
K.~{Hoffmann}, J.~Bel and E.~Gazta{\~n}aga, \emph{{Comparing halo bias from
  abundance and clustering}}, {\emph{ArXiv e-prints} (Mar., 2015) },
  [\href{https://arxiv.org/abs/1503.00313}{{\ttfamily 1503.00313}}].

\bibitem{LazeyrasEtal2016}
T.~{Lazeyras}, C.~Wagner, T.~Baldauf and F.~Schmidt, \emph{{Precision
  measurement of the local bias of dark matter halos}},
  \href{http://dx.doi.org/10.1088/1475-7516/2016/02/018}{\emph{\jcap}
  {\bfseries 2} (Feb., 2016) 018},
  [\href{https://arxiv.org/abs/1511.01096}{{\ttfamily 1511.01096}}].

\bibitem{ModiCastorinaSeljak2017}
C.~{Modi}, E.~{Castorina} and U.~{Seljak}, \emph{{Halo bias in Lagrangian
  space: estimators and theoretical predictions}},
  \href{http://dx.doi.org/10.1093/mnras/stx2148}{\emph{\mnras} {\bfseries 472}
  (Dec., 2017) 3959--3970}, [\href{https://arxiv.org/abs/1612.01621}{{\ttfamily
  1612.01621}}].

\bibitem{SanchezEtal2017b}
A.~G. {S{\'a}nchez}, R.~{Scoccimarro}, M.~{Crocce}, J.~N. {Grieb},
  S.~{Salazar-Albornoz}, C.~{Dalla Vecchia} et~al., \emph{{The clustering of
  galaxies in the completed SDSS-III Baryon Oscillation Spectroscopic Survey:
  Cosmological implications of the configuration-space clustering wedges}},
  \href{http://dx.doi.org/10.1093/mnras/stw2443}{\emph{\mnras} {\bfseries 464}
  (Jan., 2017) 1640--1658}, [\href{https://arxiv.org/abs/1607.03147}{{\ttfamily
  1607.03147}}].

\bibitem{BeutlerEtal2017}
F.~{Beutler}, H.-J. {Seo}, A.~J. {Ross}, P.~{McDonald}, S.~{Saito}, A.~S.
  {Bolton} et~al., \emph{{The clustering of galaxies in the completed SDSS-III
  Baryon Oscillation Spectroscopic Survey: baryon acoustic oscillations in the
  Fourier space}}, \href{http://dx.doi.org/10.1093/mnras/stw2373}{\emph{\mnras}
  {\bfseries 464} (Jan., 2017) 3409--3430},
  [\href{https://arxiv.org/abs/1607.03149}{{\ttfamily 1607.03149}}].

\bibitem{SlepianEisenstein2017}
Z.~{Slepian} and D.~J. {Eisenstein}, \emph{{Modelling the large-scale
  redshift-space 3-point correlation function of galaxies}},
  \href{http://dx.doi.org/10.1093/mnras/stx490}{\emph{\mnras} {\bfseries 469}
  (Aug., 2017) 2059--2076}, [\href{https://arxiv.org/abs/1607.03109}{{\ttfamily
  1607.03109}}].

\bibitem{CastorinaEtal2016}
E.~{Castorina}, A.~{Paranjape}, O.~{Hahn} and R.~K. {Sheth}, \emph{{Excursion
  set peaks: the role of shear}}, {\emph{ArXiv e-prints} (Nov., 2016) },
  [\href{https://arxiv.org/abs/1611.03619}{{\ttfamily 1611.03619}}].

\bibitem{CastorinaParanjapeSheth2017}
E.~{Castorina}, A.~{Paranjape} and R.~K. {Sheth}, \emph{{Constraints on halo
  formation from cross-correlations with correlated variables}},
  \href{http://dx.doi.org/10.1093/mnras/stx701}{\emph{\mnras} {\bfseries 468}
  (July, 2017) 3813--3827}, [\href{https://arxiv.org/abs/1611.03613}{{\ttfamily
  1611.03613}}].

\bibitem{HoffmannBelGaztanaga2017}
K.~{Hoffmann}, J.~{Bel} and E.~{Gazta{\~n}aga}, \emph{{Linear and non-linear
  bias: predictions versus measurements}},
  \href{http://dx.doi.org/10.1093/mnras/stw2876}{\emph{\mnras} {\bfseries 465}
  (Feb., 2017) 2225--2235}, [\href{https://arxiv.org/abs/1607.01024}{{\ttfamily
  1607.01024}}].

\bibitem{ChuangEtal2017}
C.-H. {Chuang}, M.~{Pellejero-Ibanez}, S.~{Rodr{\'{\i}}guez-Torres}, A.~J.
  {Ross}, G.-b. {Zhao}, Y.~{Wang} et~al., \emph{{The clustering of galaxies in
  the completed SDSS-III Baryon Oscillation Spectroscopic Survey: single-probe
  measurements from DR12 galaxy clustering - towards an accurate model}},
  \href{http://dx.doi.org/10.1093/mnras/stx1641}{\emph{\mnras} {\bfseries 471}
  (Oct., 2017) 2370--2390}, [\href{https://arxiv.org/abs/1607.03151}{{\ttfamily
  1607.03151}}].

\bibitem{Wang+14}
L.~{Wang}, B.~{Reid} and M.~{White}, \emph{{An analytic model for
  redshift-space distortions}},
  \href{http://dx.doi.org/10.1093/mnras/stt1916}{\emph{\mnras} {\bfseries 437}
  (Jan., 2014) 588--599}, [\href{https://arxiv.org/abs/1306.1804}{{\ttfamily
  1306.1804}}].

\bibitem{VlahSeljakBaldauf2015}
Z.~{Vlah}, U.~{Seljak} and T.~{Baldauf}, \emph{{Lagrangian perturbation theory
  at one loop order: Successes, failures, and improvements}},
  \href{http://dx.doi.org/10.1103/PhysRevD.91.023508}{\emph{\prd} {\bfseries
  91} (Jan., 2015) 023508}, [\href{https://arxiv.org/abs/1410.1617}{{\ttfamily
  1410.1617}}].

\end{thebibliography}\endgroup

\end{document}